\documentclass[%
 reprint,
superscriptaddress,
 amsmath,amssymb,
 aps,
]{revtex4-1}

\usepackage{graphicx}
\usepackage{dcolumn}
\usepackage{bm}
\usepackage{amsmath} 
\renewcommand{\vec}[1]{\boldsymbol{#1}} 
\usepackage{verbatim}

\begin{document}


\title{Localization, phases and transitions in the three-dimensional extended Lieb lattices}

\author{Jie Liu}
\email{liujie@smail.xtu.edu.cn}
\affiliation{School of Physics and Optoelectronics, Xiangtan University, Xiangtan 411105, China}

\author{Xiaoyu Mao}
\email{Maoxiaoyu@smail.xtu.edu.cn}
\affiliation{School of Physics and Optoelectronics, Xiangtan University, Xiangtan 411105, China}

\author{Jianxin Zhong}
\email{jxzhong@xtu.edu.cn}
\affiliation{School of Physics and Optoelectronics, Xiangtan University, Xiangtan 411105, China}

\author{Rudolf A. R\"omer}
\affiliation{School of Physics and Optoelectronics, Xiangtan University, Xiangtan 411105, China}
\email{r.roemer@warwick.ac.uk}
\affiliation{Department of Physics, University of Warwick, Coventry, CV4 7AL,
	United Kingdom}
\affiliation{
CY Advanced Studied and LPTM (UMR8089 of CNRS),
CY Cergy-Paris Universit\'{e},
F-95302 Cergy-Pontoise, France%
}

\date{\today}

\begin{abstract}
We study the localization properties and the Anderson transition in the 3D Lieb lattice $\mathcal{L}_3(1)$ and its extensions $\mathcal{L}_3(n)$ in the presence of disorder. We compute the positions of the flat bands, the disorder-broadened density of states and the energy-disorder phase diagrams for up to 4 different such Lieb lattices. Via finite-size scaling, we obtain the critical properties such as critical disorders and energies as well as the universal localization lengths exponent $\nu$. We find that the critical disorder $W_c$ decreases from $\sim 16.5$ for the cubic lattice, to $\sim 8.6$ for $\mathcal{L}_3(1)$, $\sim 5.9$ for $\mathcal{L}_3(2)$ and $\sim 4.8$ for $\mathcal{L}_3(3)$. Nevertheless, the value of the critical exponent $\nu$ for all Lieb lattices studied here and across disorder and energy transitions agrees within error bars with the generally accepted universal value $\nu=1.590  (1.579,1.602)$. 
\end{abstract}

\maketitle


\section{\label{sec:intro}Introduction}


Flat energy bands have recently received renewed attention due to much experimental progress in the last decade \cite{Leykam2018Perspective:Flatbands}. The hallmark of such flat bands is an absence of dispersion in the whole of $k$-space \cite{Tasaki1998FromModel,Miyahara2007BCSLattice,Bergman2008BandModels,Wu2007FlatLattice}, implying an effectively zero kinetic energy. This leads to a whole host of effects in transport and optical response such as, e.g.\ localization of eigenstates without disorder \cite{Leykam2017LocalizationStates} and enhanced optical absorption and radiation. Further studies explorations of flat-band physics have now been done in Wigner crystals \cite{Wu2007FlatLattice}, high-temperature superconductors \cite{Miyahara2007BCSLattice,Julku2016GeometricBand}, photonic wave guide arrays \cite{Vicencio2015a,Mukherjee2015a,Guzman-Silva2014ExperimentalLattices,Diebel2016ConicalLattices,Leykam2018Perspective:Flatbands}, Bose-Einstein  condensates \cite{Baboux2016BosonicBand,Taie2015CoherentLattice}, ultra-cold atoms in optical lattices \cite{Shen2010SingleLattices} and electronic systems \cite{Slot2017ExperimentalLattice}.

Systems that exhibit flat-band physics correspond usually to 
specially "engineered" lattice structures such as quasi-1D lattices \cite{Leykam2017LocalizationStates,Shukla2018,Ramachandran2017}, diamond-type lattices \cite{Goda2006InverseFlatbands}, and so-called Lieb lattices,\cite{Julku2016GeometricBand,Qiu2016DesigningSurface,Chen2017Disorder-inducedLattices,Nita2013SpectralLattice,Sun2018ExcitationLattice,Bhattacharya2019a}. 
%
Indeed, the Lieb lattice, a two-dimensional (2D) extension of a simple cubic lattice, was the first where the flat band structure was recognized and used to enhance magnetic effects in model studies \cite{Lieb1989TwoModel,Mielke1993FerromagnetismModel,Tasaki1998FromModel}. Most other flat-band systems cited above are also of the Lieb type and exists as either 2D, quasi-1D or 1D lattices \cite{Lee2019HiddenStructures}.
%
Less attention has been given to 3D flat-band systems \cite{Goda2006InverseFlatbands} or extended Lieb lattices \cite{Bhattacharya2019a,Mao2020}. Furthermore, until recently hardly any work has investigated the influence of disorder on flat-band systems \cite{Shukla2018,Shukla2018a}. Recently, instead of concentrating on the  properties of flat-band states, we investigated how the localization properties in the neighboring dispersive bands are changed by the disorder for 2D flat-band systems \cite{Mao2020}. 

In the present work, we extend these studies to the class of 3D extended Lieb lattices. As is well known \cite{Krameri1993} the Anderson transition in a simple cubic lattice with uniform potential disorder $\epsilon_{\vec{x}} \in [-W/2,W/2] $ at each site $\vec{x}$ is characterized by a critical disorder $W_c=16.0(5) t$ \cite{MacKinnon1981One-ParameterSystems}, with $t$ denoting the nearest neighbor hopping strength. The full energy-disorder phase diagram is characterized by a simple-connected region of extended states ranging from $\pm 6t$ at $W=0$ and ending at $W_c=16.530(16.524,16.536)$ for $E=0$ \cite{Rodriguez2011MultifractalTransition}. The critical exponent of the transition has been determined with ever greater precision as close to, e.g., $\nu= 1.590 (1.579,1.602)$ \cite{Rodriguez2011MultifractalTransition} and $1.57(2)$ \cite{Slevin1999b}.
%
The 3D Lieb model, shown in Fig.\ \ref{Fig:Lieb3D_Graph} together with its extensions, is characterized by additional sites on the edges between the original site of the cubic lattice. As such, the transport along the edges should become more 1D-like and we expect that the phase diagram should have a smaller region of extended states. 

\section{\label{sec:model}Models and Method}

\subsection{\label{sec:tmm}\label{sec:lieb}Transfer-matrix method for the 3D Lieb lattices and its extensions $\mathcal{L}_3(n)$}

\begin{figure}[tbh]
    \centering
    (a)\includegraphics[width=0.27\columnwidth]{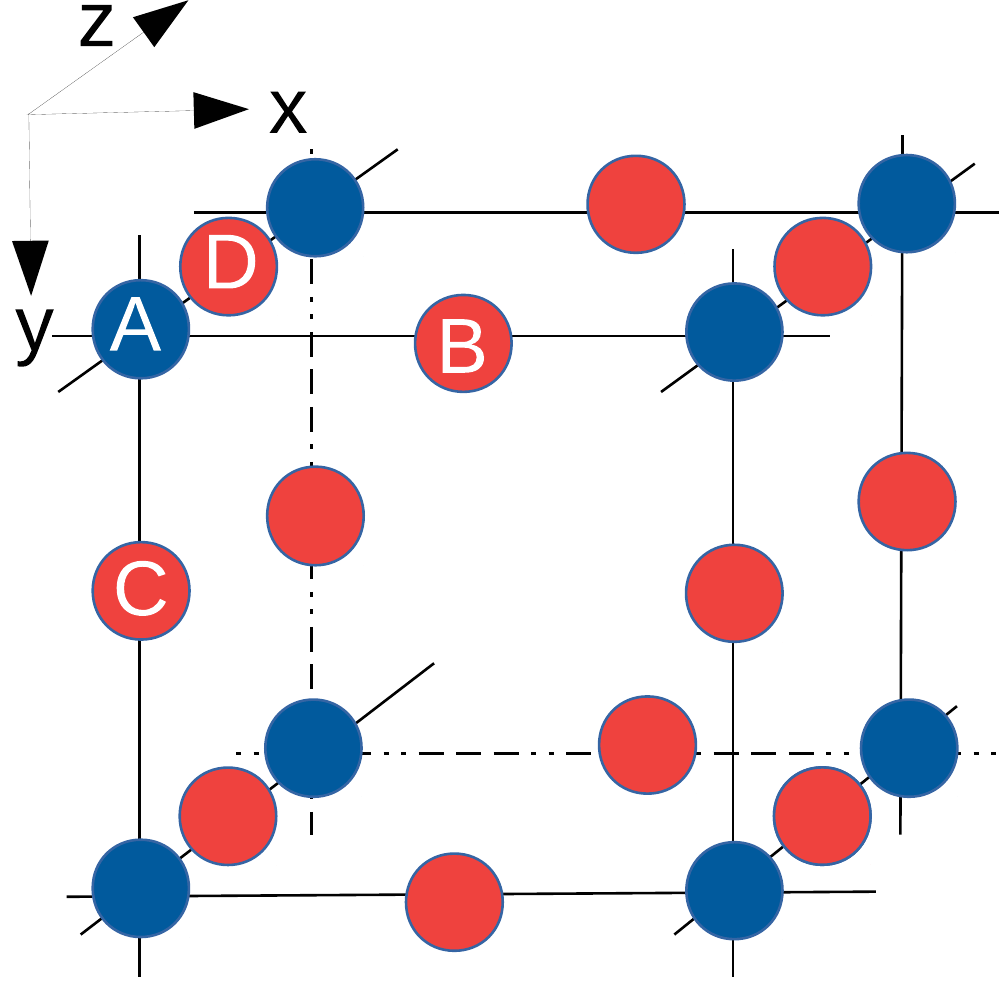}
    (b)\includegraphics[width=0.27\columnwidth]{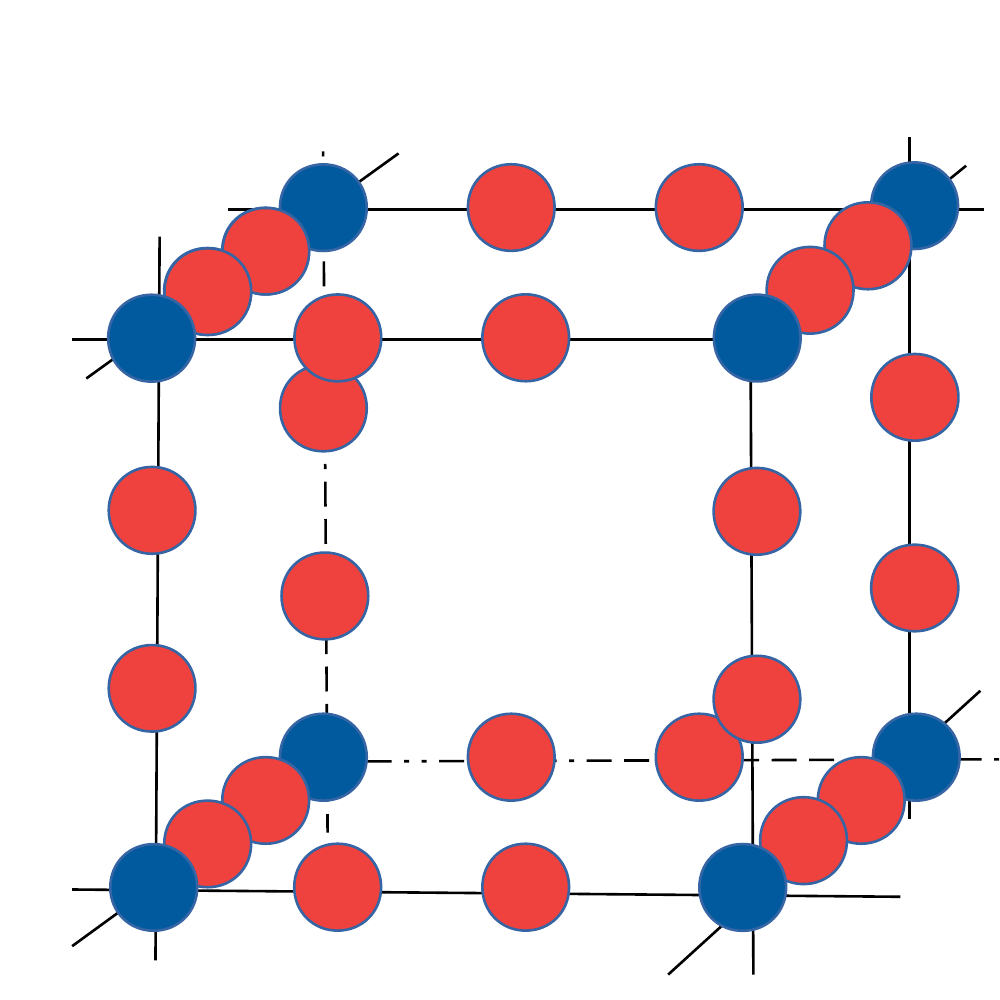}
    (c)\includegraphics[width=0.27\columnwidth]{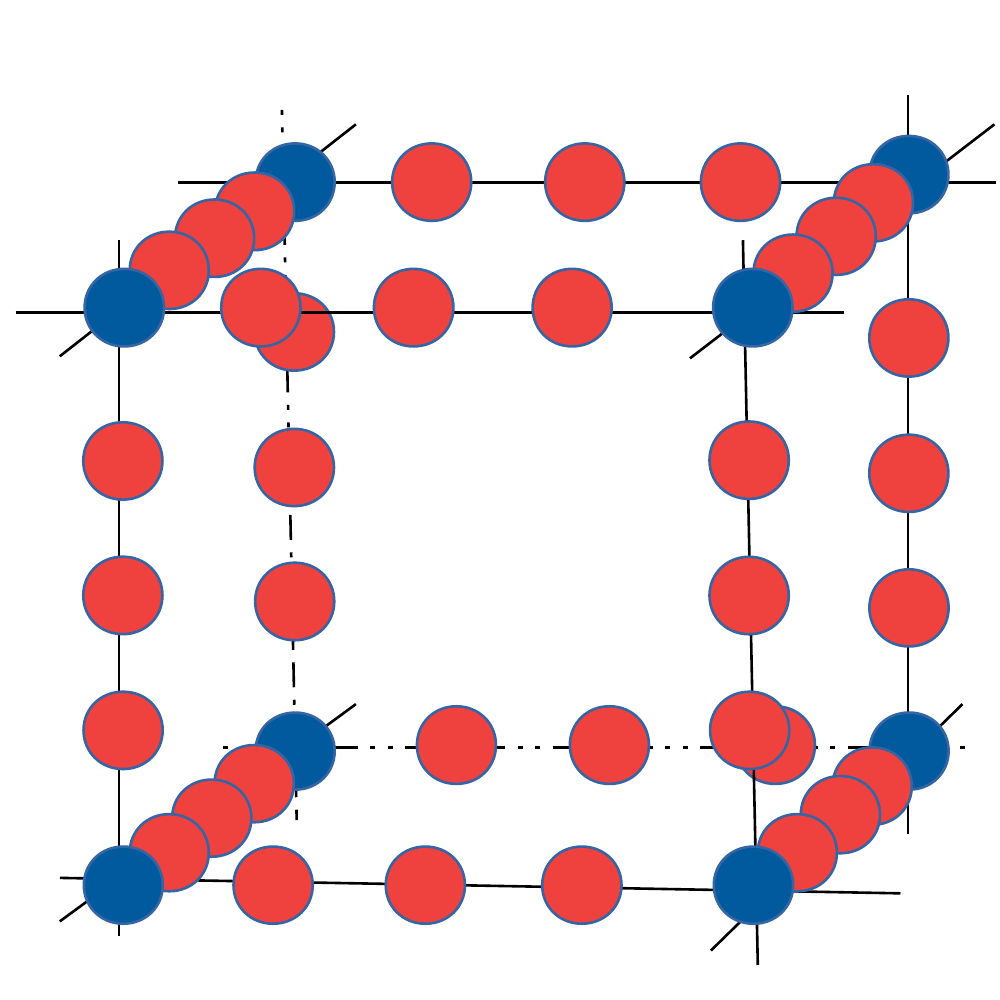}
    \caption{(a) The Lieb lattice $\mathcal{L}_3(1)$ and its extensions (b) $\mathcal{L}_3(2)$ and (c) $\mathcal{L}_3(3)$. The blue spheres denote the original nearest-neighbor sites in the underlying cubic lattice while the red spheres show the added sites. The solid lines indicate the cubic structure. The coordinate system is to help identify the TMM setup used in our study as are the labels A, B, C, and D.
    }
    \label{Fig:Lieb3D_Graph}
\end{figure}

We denote the Lieb lattices as $\mathcal{L}_d(n)$ if there are $n$ equally-spaced atoms between two original nearest neighbors in a $d$-dimensional lattice. Here, we shall concentrate on $\mathcal{L}_3(1)$, $\mathcal{L}_3(2)$ and $\mathcal{L}_3(3)$ as shown in Fig.\ \ref{Fig:Lieb3D_Graph}.   
To explore the effects of disorder, we use the standard Anderson Hamiltonian 
\begin{equation}
    H=\sum_{\vec{x}} \epsilon_{\vec{x}} |\vec{x}\rangle \langle \vec{x}|
    - \sum_{\vec{x}\neq \vec{y}}t_{\vec{x}\vec{y}}|\vec{x}\rangle \langle \vec{y}|
\label{Equ:def1}. 
\end{equation}
The orthonormal Wannier states $|\vec{x}\rangle$ describes electrons located at sites $\vec{x}=(x,y,z)$ of Lieb lattice with hard boundary condition (we have similar results for periodic boundary conditions as well).
The hopping integrals $t_{\vec{x}\vec{y}}= t$ only for $\vec{x}$, $\vec{y}$ being nearest neighbors as indicated by the lines in Fig.\ \ref{Fig:Lieb3D_Graph}, otherwise $t_{\vec{x}\vec{y}}= 0$. 

For $\mathcal{L}_3(1)$, in order to calculate the localization length $\lambda$ of the wave function by the transfer-matrix method (TMM), we consider a quasi-one-dimensional bar, with cross area $M^2$ and length $ L>>M $. A unit length corresponds to original site-to-site distances as indicated by A sites in Fig.\ \ref{Fig:Lieb3D_Graph}. Along the transfer axis in the $z$-direction, there are two different slices in $\mathcal{L}_3(1)$. The first slice contains the original A sites, and the added B and C sites to form an A-B-C slice, the second (D-)slice only contains the added D sites as shown in Fig.\ \ref{Fig:Lieb3D_Graph}. The TMM equation implementing $H \Psi=E \Psi$ at energy $E$ for the Hamiltonian \eqref{Equ:def1} can be written as two parts. First, transferring from slice A-B-C to slice D slice, we have
\begin{widetext}
\begin{equation}
\begin{aligned}
\left( 
\begin{array}{c}
    \Psi_{z+1}^{D}\\
    \\
    \Psi_{z}^{A}
\end{array}
\right) &= 
\mathbf{T}_{A\to D}
\left( 
\begin{array}{c}
\Psi_{z}^{A}\\
\\
\Psi_{z-1}^{D}
\end{array}
\right)\\
&=\left( 
\begin{array}{cc}
  \mathcal{E} \mathbf{1}_{M^2}
 -\frac{1}{\epsilon_{z,x-1,y}-E} \mathbf{t}_{x-}
 -\frac{1}{\epsilon_{z,x+1,y}-E} \mathbf{t}_{x+}
 -\frac{1}{\epsilon_{z,x,y-1}-E} \mathbf{t}_{y-}
 -\frac{1}{\epsilon_{z,x,y+1}-E} \mathbf{t}_{y+}&
  -\mathbf{1}_{M^2}\\
\\
\mathbf{1}_{M^2}&    \mathbf{0}_{M^2}
\end{array}
\right)
\left( 
    \begin{array}{c}
    \Psi_{z}^{A}\\
    \\
    \Psi_{z-1}^{D}
\end{array}
\right),
\label{eq:tmm}
\end{aligned}
\end{equation}
where
\begin{equation}
\begin{aligned}
  \mathcal{E} &= \frac{\epsilon_{z,x,y}-E}{t}-\frac{t}{\epsilon_{z,x-1,y}-E}
  -\frac{t}{\epsilon_{z,x+1,y}-E}
  -\frac{t}{\epsilon_{z,x,y-1}-E}
  -\frac{t}{\epsilon_{z,x,y+1}-E} ,
\end{aligned}
\end{equation}
\end{widetext}
and $\mathbf{0}_{M^2}$,  $\mathbf{1}_{M^2}$ denote $M^2 \times M^2$ zero and identity matrices, respectively. Similarly, $ \textbf{t}_{x+} $, $ \textbf{t}_{x-} $, $ \textbf{t}_{y+} $ and $ \textbf{t}_{y-} $ are $M^2 \times M^2$ connectivity matrices in the positive/negative $x$/$y$ directions. With this choice of TMM set-up, we effectively renormalize the added B, C (red) sites shown in Fig.\ \ref{Fig:Lieb3D_Graph}(a). Taking $M=3$ as an example, we can explicitely write the $3^2 \times 3^2$ matrices
\begin{equation}
\begin{aligned}
\textbf{t}_{x-} =& t \left(
     \begin{array}{ccccccccccc}
        0&      1&      0&      0&      0&      0&      0&      0&      0&    \\
        0&      0&      1&      0&      0&      0&      0&      0&      0&    \\
      (1)&      0&      0&      0&      0&      0&      0&      0&      0&    \\
        0&      0&      0&      0&      1&      0&      0&      0&      0&    \\
        0&      0&      0&      0&      0&      1&      0&      0&      0&    \\
        0&      0&      0&    (1)&      0&      0&      0&      0&      0&    \\
        0&      0&      0&      0&      0&      0&      0&      1&      0&    \\
        0&      0&      0&      0&      0&      0&      0&      0&      1&    \\
        0&      0&      0&      0&      0&      0&    (1)&      0&      0&    
    \end{array}
\right)   
\label{eq:tx-}
 \end{aligned}
\end{equation}
and $\textbf{t}_{x+} =  \textbf{t}_{x-}^{\dag}$. Similarly,
\begin{equation}
\begin{aligned}
\textbf{t}_{y-} =&t \left(
     \begin{array}{cccccccccc}
        0&      0&      0&      1&      0&      0&      0&      0&      0&    \\
        0&      0&      0&      0&      1&      0&      0&      0&      0&    \\
        0&      0&      0&      0&      0&      1&      0&      0&      0&    \\
        0&      0&      0&      0&      0&      0&      1&      0&      0&    \\
        0&      0&      0&      0&      0&      0&      0&      1&      0&    \\
        0&      0&      0&      0&      0&      0&      0&      0&      1&    \\
      (1)&      0&      0&      0&      0&      0&      0&      0&      0&    \\
        0&    (1)&      0&      0&      0&      0&      0&      0&      0&    \\
        0&      0&    (1)&      0&      0&      0&      0&      0&      0& 
    \end{array}
\right)   
\label{eq:ty-}
 \end{aligned}
\end{equation}
and $\textbf{t}_{y+} = \textbf{t}_{y-}^{\dag}$. In Eqs.\ \eqref{eq:tx-} and \eqref{eq:ty-}, the matrix entries $(1)$ can be chosen $0$ for hard-wall boundaries and $1$ for periodic boundaries. In this way, the effects of sites B and C have been renormalized into effective onsite energies $\mathcal{E}$ and hopping terms $\textbf{t}_{x\pm}$, $\textbf{t}_{y\pm}$ keeping the transfer matrix $\mathbf{T}_{A\rightarrow D}$ in the standard $2 M^2 \times 2 M^2$ form. 
We emphasize that $\Psi^{A,D}_{z}$ denotes a vector of length $M^2$ for wave function amplitudes in the $z$th slice \cite{psiAD}, either A or D, with $x, y=1, \ldots, M$, labelling the position of the original cubic sites in this slice. 
In this notation the term $\frac{\epsilon_{z,x,y}-E}{t} \mathbf{1}_{M^2} \equiv \mathrm{diag}\left(\frac{\epsilon_{z,1,1}-E}{t},\frac{\epsilon_{z,1,2}-E}{t}, \ldots, \frac{\epsilon_{z,M,M}-E}{t}\right)$ and similarly for $\mathcal{E}\mathbf{1}_{M^2}$ and the hopping terms with $\textbf{t}_{x\pm}$, $\textbf{t}_{y\pm}$ in Eq.\ \eqref{eq:tmm}.
From the D slice to the A-B-C slice, we can write a more standard TMM form as
\begin{equation}
\begin{aligned}
\left( 
    \begin{array}{*{20}{c}}
        \Psi_{z+1}^{A}\\ 
        \\
        \Psi_{z}^{D}
    \end{array}
    \right)
 &=\mathbf{T}_{D\to A} 
 \left( 
    \begin{array}{*{20}{c}}
       \Psi _{z}^{D} \\
       \\
       \Psi _{z-1}^{A}
    \end{array}
\right) \\
&=\left( 
\begin{array}{*{20}{c}}
  \frac{\epsilon_{z,x,y}-E}{t} \mathbf{1}_{M^2}&     -\mathbf{1}_{M^2}\\
  \\
  \mathbf{1}_{M^2}&                                   \mathbf{0}_{M^2}
\end{array}
\right)
\left( 
\begin{array}{*{20}{c}}
\Psi _{z}^{D}\\
\\
\Psi _{z-1}^{A}
\end{array} 
\right)    .
\end{aligned}
\end{equation}
in similar notation.

The TMM method proceeds by multiplying successively $\mathbf{T}_{A\to D}$ by $\mathbf{T}_{D\to A}$ along the bar in $z$-direction, using $M^2$ possible starting vector $\Psi_z^A(1)=(1, 0, \ldots, 0)$, $\Psi_z^A(2)=(0, 1, \ldots, 0)$, $\Psi_z^A(M^2)=(0, 0, \ldots, 1)$ to form a complete set. We regularly renorthogonalize these $M^2$ $\Psi$ states, usually after every 10th multiplication. The Lyapunov exponents $\gamma_i$, $i=1, 2, \dots, M^2$, and their accumulated changes are calculated until a preset precision is reached for the smallest $\gamma_\mathrm{min}$ \cite{Krameri1993,Oseledets1968ASystems,Ishii1973LocalizationSystem,Beenakker1997Random-matrixTransport}. The localization length $\lambda(M,E,W)=\gamma_\mathrm{min} >0$, the dimensionless reduced localization length is $\Lambda_M(E,W)=\lambda(M,E,W)/M$.
These considerations set out the TMM for $\mathcal{L}_3(1)$. For the extended Lieb lattices, we follow a similar strategy, leading to an even more involved renormalization scheme which we refrain to review in the interest of brevity.
   
\subsection{\label{sec:FSS} Finite-size scaling}

The metal-insulator transition (MIT) in the Anderson model of localization is expected to be a second-order phase transition, characterized by a divergence in a correlation length $\xi(W)\propto\left|W-W_{c}\right|^{-\nu}$ at fixed energy $E$, and $\xi(E)\propto\left|E-E_{c}\right|^{-\nu}$ at fixed disorder $W$ \cite{EILMES2008}, where $E_{c}$ is the critical energy and $\nu$, $W_{c}$ as before. 

We determine the reduced correlation length $\xi/M$ in the thermodynamic limit assuming the single parameter scaling i.e.\  
$
 \Lambda_{M}(M,E,W)=f(\xi/M)
$ \cite{MacKinnon1981One-ParameterSystems}. 
For a system with an MIT this scaling function consists of two branches corresponding to localized and extended phases. 
Using finite-size scaling (FSS) \cite{MacKinnon1983a}, we can obtain estimates of the critical exponent.
Here, we use a method \cite{EILMES2008,Slevin1999b} that models two kinds of corrections to scaling: (i) the presence of irrelevant scaling variables and (ii) non-linearity of the scaling variables. Hence one writes 
$
    \Lambda=F(\chi_{r}M^{1/\nu},\chi_{i}M^{y})
    \label{eq1}.
$, 
where $\chi_{r}$ the relevant scaling variable and $\chi_{i}$ the irrelevant scaling variable. 
We next Taylor-expand $\Lambda$ and $F$ up to order $n_{i}$ and $n_{r}$ such that 
\begin{eqnarray}
    \Lambda {\rm{ = }}\sum\limits_{n = 0}^{{n_i}} {\chi _i^n{M^{ny}}{F_n}({\chi _r}{M^{1/\upsilon }})}
    \label{eq19},
    {F_n}=\sum\limits_{k=0}^{{n_r}}a_{nk} {\chi _r^k} {M^{k/\nu}}
    \label{eq20}.
\end{eqnarray}
Furthermore, we also expand $\chi_{i}$ and $\chi_{r}$ by $\omega=(W_{c}-W)/W_{c}$ (or $(E_{c}-E)/E_{c}$) to consider the importance of the nonlinearities, 
\begin{equation}
    {\chi _r}(\omega ) = \sum\limits_{m = 1}^{{m_r}} {{b_m}{\omega ^m}},
    {\chi _i}(\omega ) = \sum\limits_{m = 0}^{{m_i}} {{c_m}{\omega ^m}}.
    \label{eq2}
\end{equation}
In order to fix the absolute scales of $\Lambda$ in \eqref{eq19} we set $b_1 = c_0 = 1$. We then perform the FSS procedure for various values of $n_i, n_r, m_i, m_r$, in order to obtain the best stable and robust fit by minimizing the $\chi^2$ statistic. We quote goodness of fit $p$ values to allow the reader to judge the quality of our results.

\section{\label{sec:results}Results}

\subsection{\label{sec:l31+2+3+4}Dispersion and disorder-broadened density of states for $ \mathcal{L}_3(n)$}

\begin{figure*}[tbh]
    \centering
     \quad\quad \quad \quad \quad \mbox{ }\quad $\mathcal{L}_3(1)$ \hfill $\mathcal{L}_3(2)$ \hfill $\mathcal{L}_3(3)$ \hfill $\mathcal{L}_3(4)$ \quad \quad \quad \mbox{ }\quad\quad \\
    (a)\includegraphics[width=0.46\columnwidth]{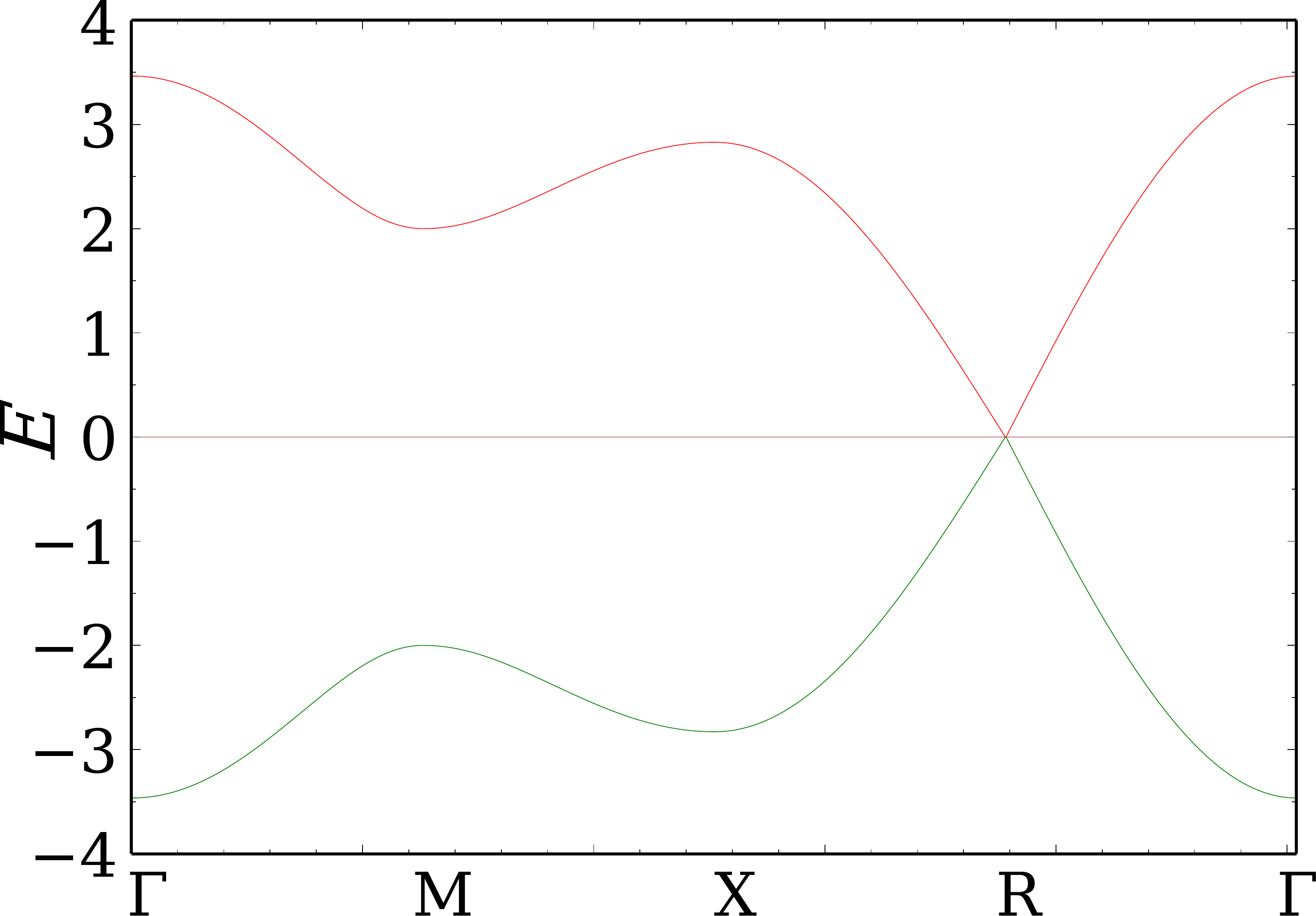}
    (b)\includegraphics[width=0.46\columnwidth]{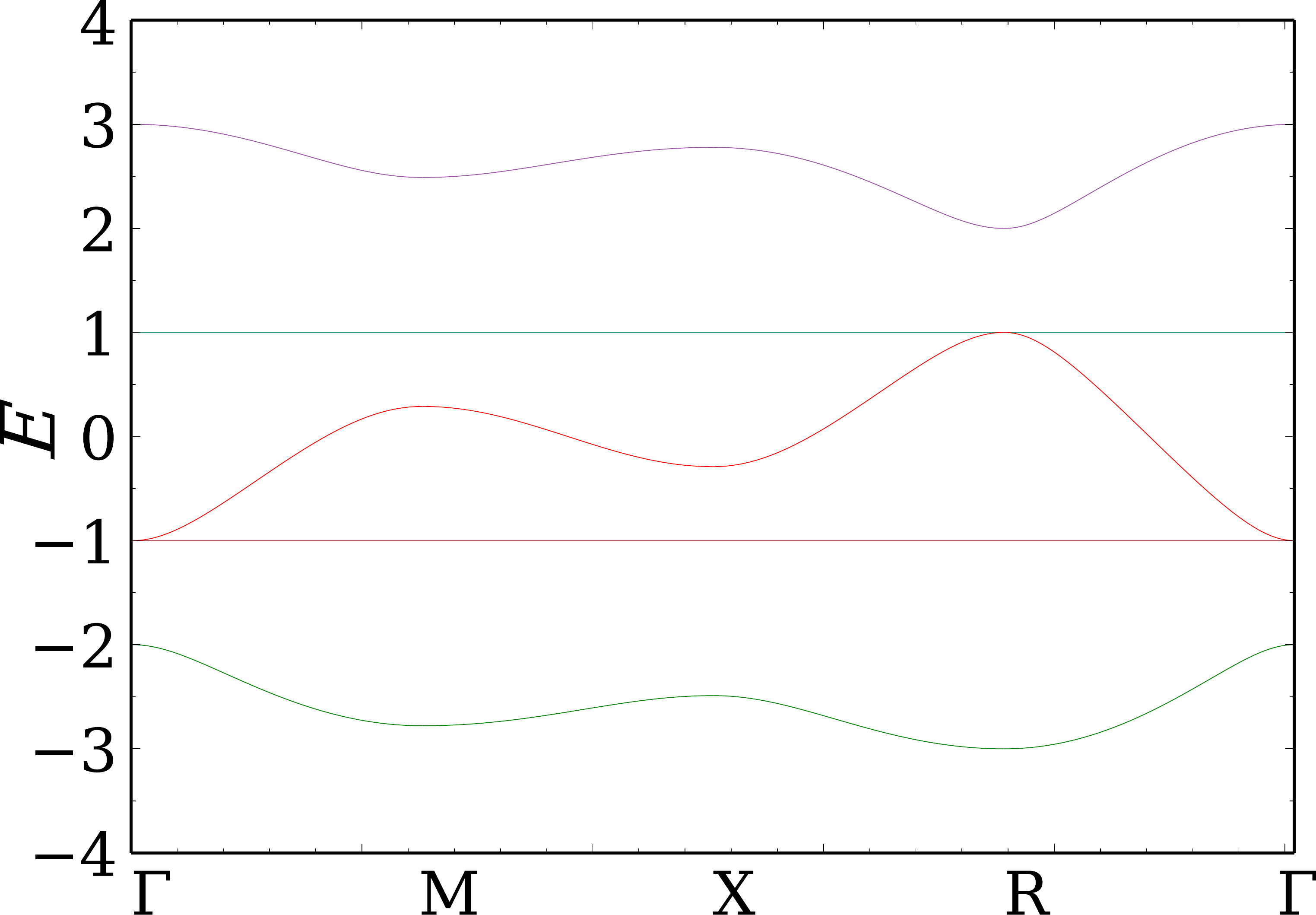}
    (c)\includegraphics[width=0.46\columnwidth]{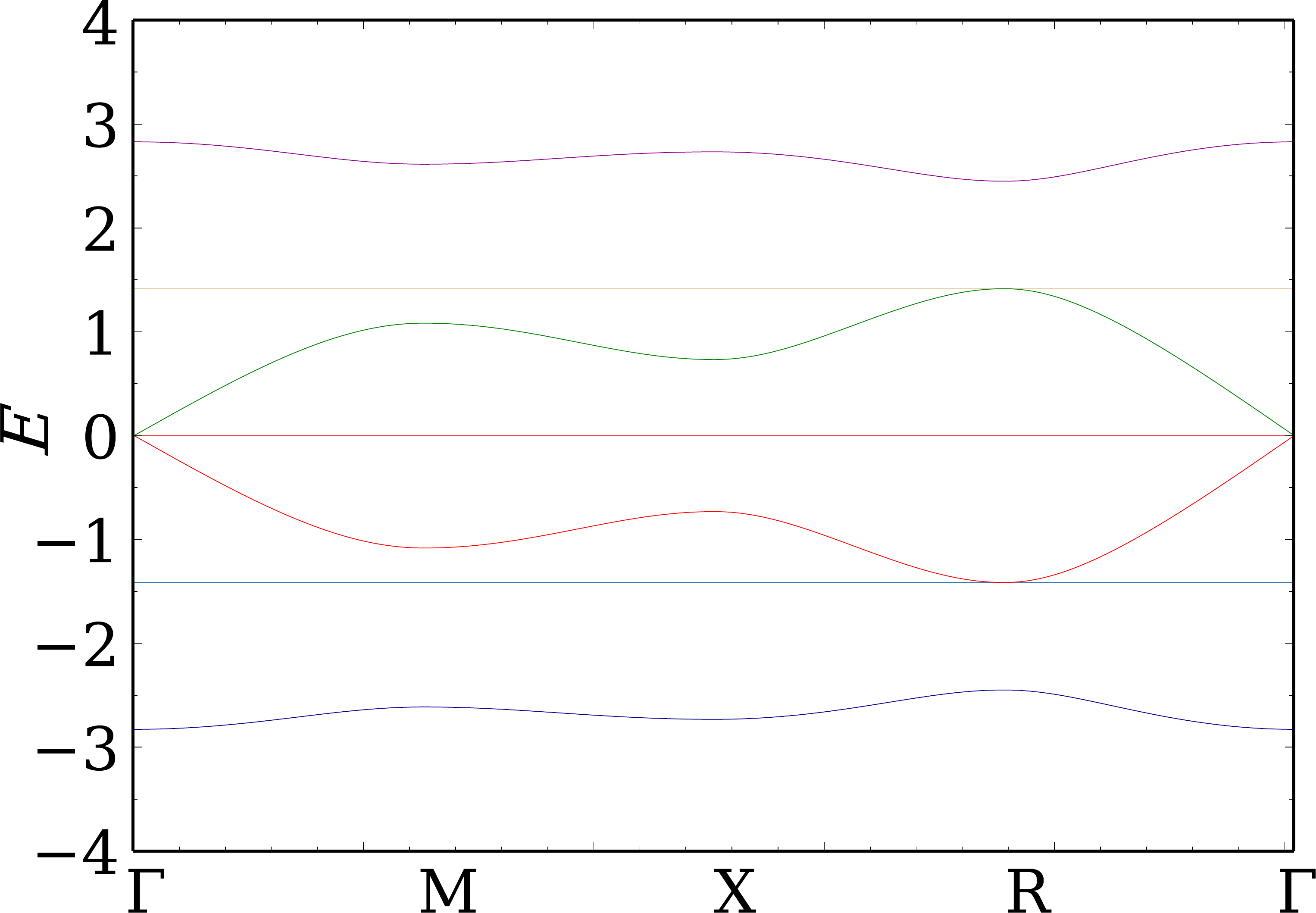}
    (d)\includegraphics[width=0.46\columnwidth]{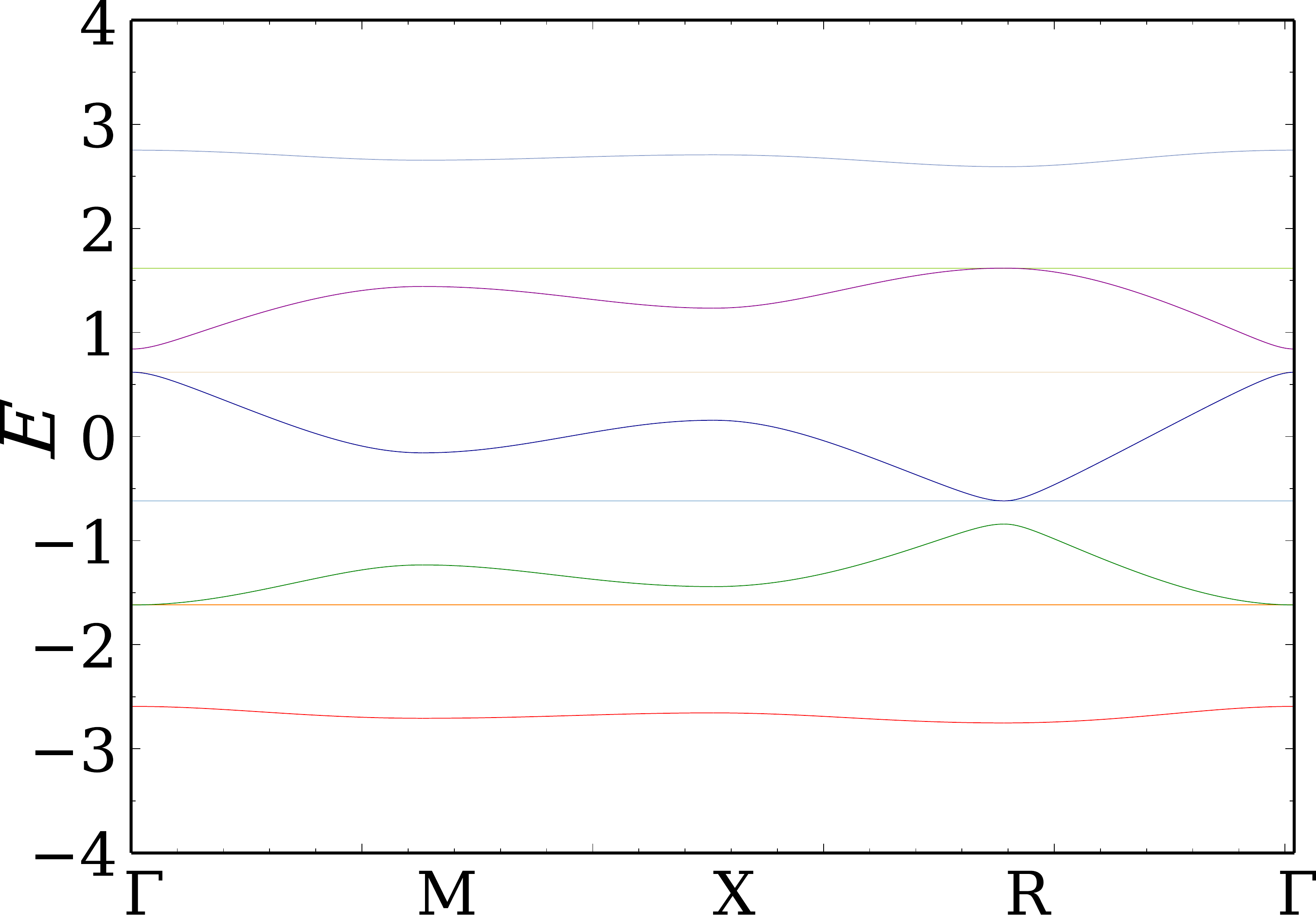}
    (e)\includegraphics[width=0.46\columnwidth]{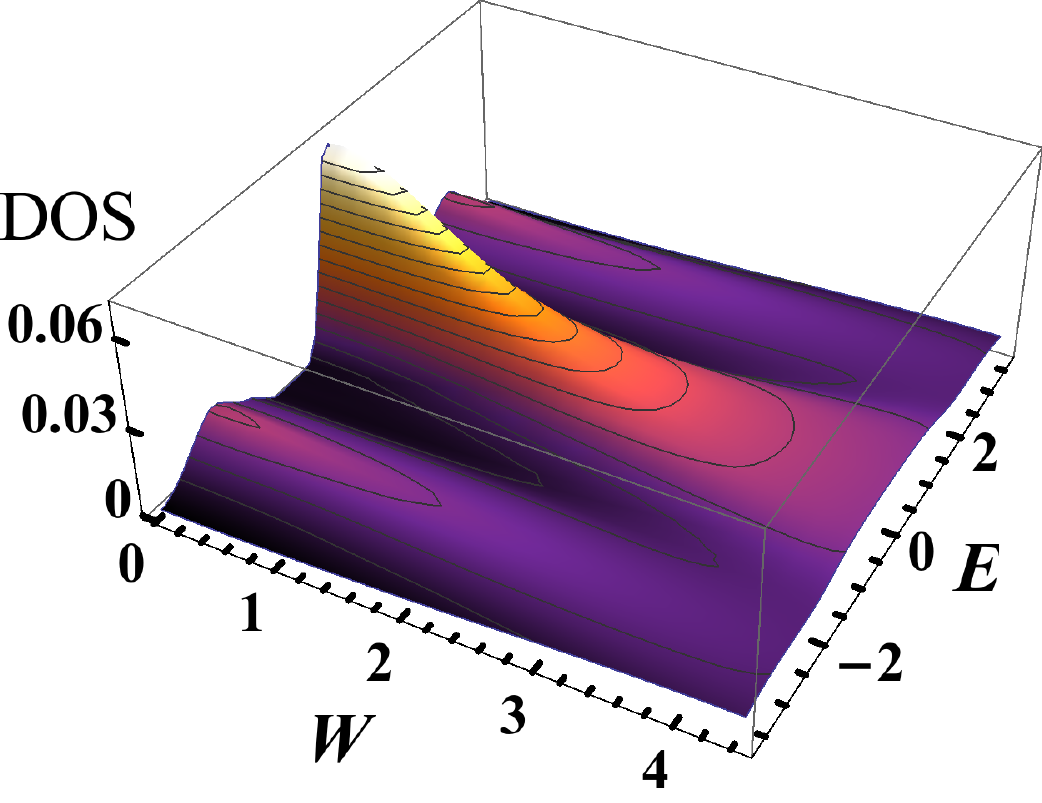}
    (f)\includegraphics[width=0.46\columnwidth]{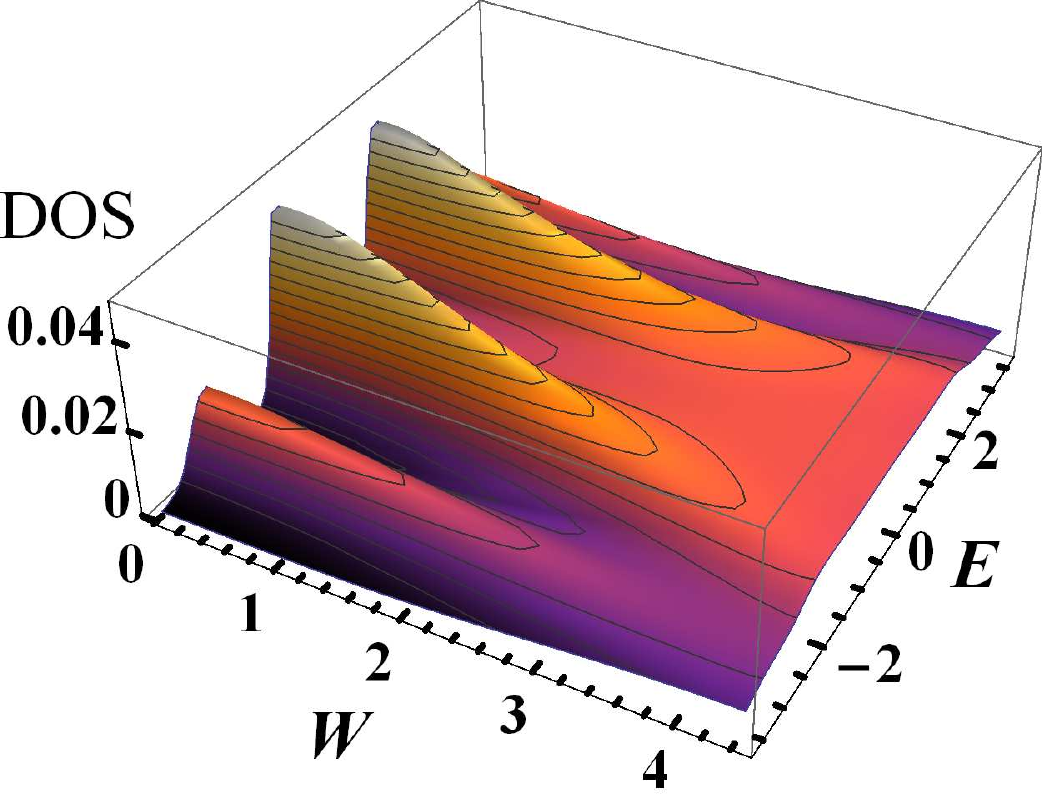}
    (g)\includegraphics[width=0.46\columnwidth]{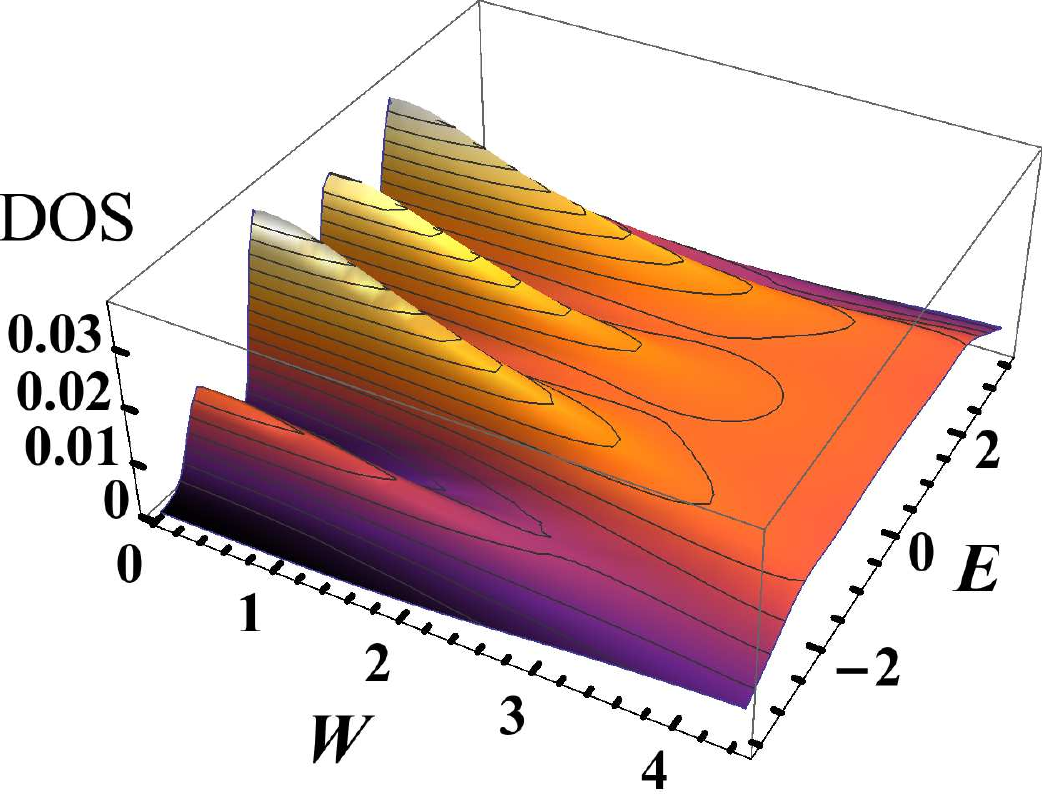}
    (h)\includegraphics[width=0.46\columnwidth]{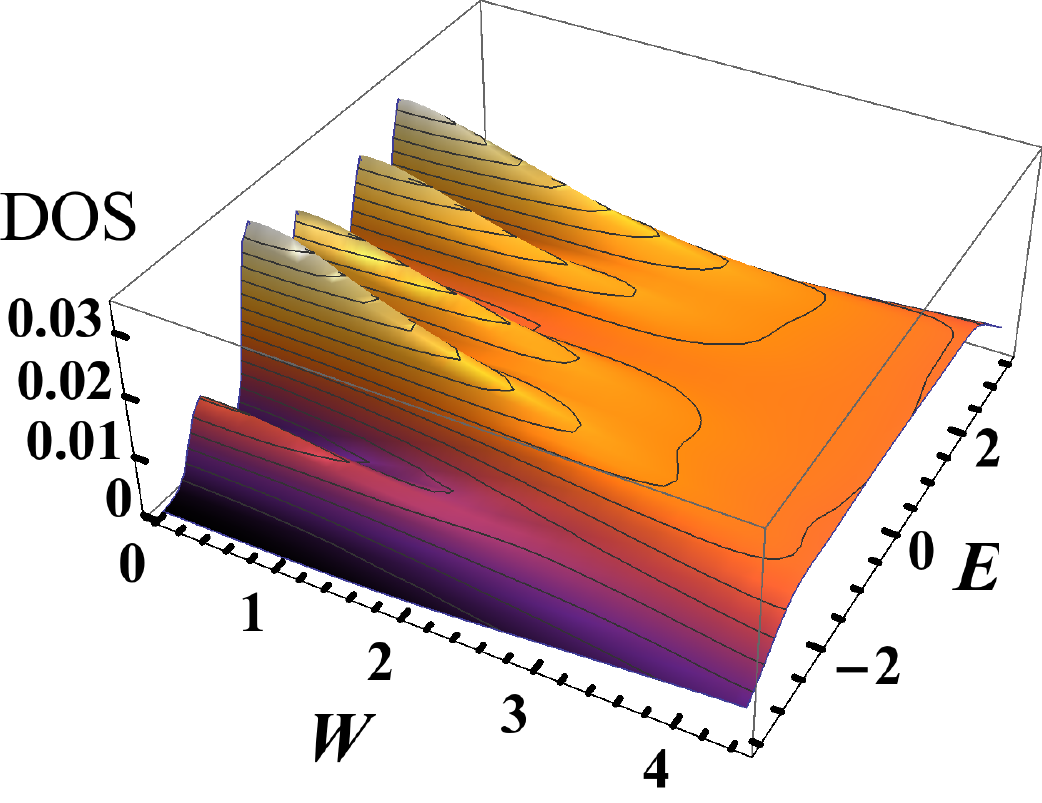}
    \caption{(a)--(d) Dispersion relations for clean systems and (e)--(h) dependence of the normalized DOS on $W$ for $\mathcal{L}_3(1)$ to $\mathcal{L}_3(4)$. In all cases, the flat bands are doubly degenerate. Different colors in the dispersion relations denote different bands while the colors in the DOS indicate different DOS values as also emphasized by the contour lines.}
    \label{fig:EnergyStructure}
\end{figure*}

For a clean $\mathcal{L}_3(1)$ system, the dispersion relation can be derived from  \eqref{Equ:def1} as 
\begin{equation}
    E_{1,2}= 0, \quad
    E_{3,4}=\pm \sqrt{6+2\left(\cos k_x +\cos k_y +\cos k_z \right)},
\end{equation}
where the $k_x, k_y, k_z$ are the reciprocal vectors corresponding to the $x$, $y$ and $z$ axes, respectively. Fig.\ \ref{fig:EnergyStructure}(a) shows the energy structure of $\mathcal{L}_3(1)$, where we can see two dispersive bands which meet linearly at the $R$ point $(k_x,k_y,k_z)=(\pi,\pi,\pi)$ at $E=0$. This coincides in energy with the doubly-degenerate flat band. Analogously, we calculate the energy structures for $\mathcal{L}_3(n)$, $n=2, 3, 4$ and plots them in Figs.\ \ref{fig:EnergyStructure}(b), (c) and (d), respectively. We can see that each $\mathcal{L}_3(n)$ lattice has $n$ doubly degenerate flat bands separating $n+1$ dispersive bands. Furthermore, the two dispersive bands at high and low energies are separated by energy gaps for these models.
We also note that for $\mathcal{L}_3(3)$ two dispersive bands again meet linearly, as for $\mathcal{L}_3(1)$, but in this instance at the $\Gamma$ point $(k_x,k_y,k_z)=(0,0,0)$ at $E=0$. No such linear behaviour can be found for $\mathcal{L}_3(n)$ with $n$ even.  

We now include the disorder, i.e.\ $W> 0$, and we calculate the disorder-dependent density of states (DOS) by direct diagonalization for small system sizes $M^3= 5^3, 5^3, 4^3$ and $4^3$ for $\mathcal{L}_3(n)$, $n=1, 2, 3, 4$, respectively. The DOS is generated from $W=0$ to $W=5.2$ in step of $0.05$ with $300$ samples for $\mathcal{L}_3(n)$, $n=1, 2, 3$, while we have $100$ samples for $\mathcal{L}_3(4)$. We also apply a Gaussian broadening of the energy levels to obtain a smoother DOS. The results are shown in Fig.\ \ref{fig:EnergyStructure}.   
For weak disorders we can clearly identify the large peaks in the DOS with the flat bands for all $\mathcal{L}_3(n)$ models. From $W \sim 3$ onward, the various peaks have merged into one broad DOS. Also, the energy gaps for $\mathcal{L}_3(n)$, $n=1, 2, 3, 4$, vanish quickly with increasing $W$.

\subsection{\label{sec:l31+2+3}Phase Diagrams}

\begin{figure}[tb]
    \centering
    \includegraphics[width=0.95\columnwidth]{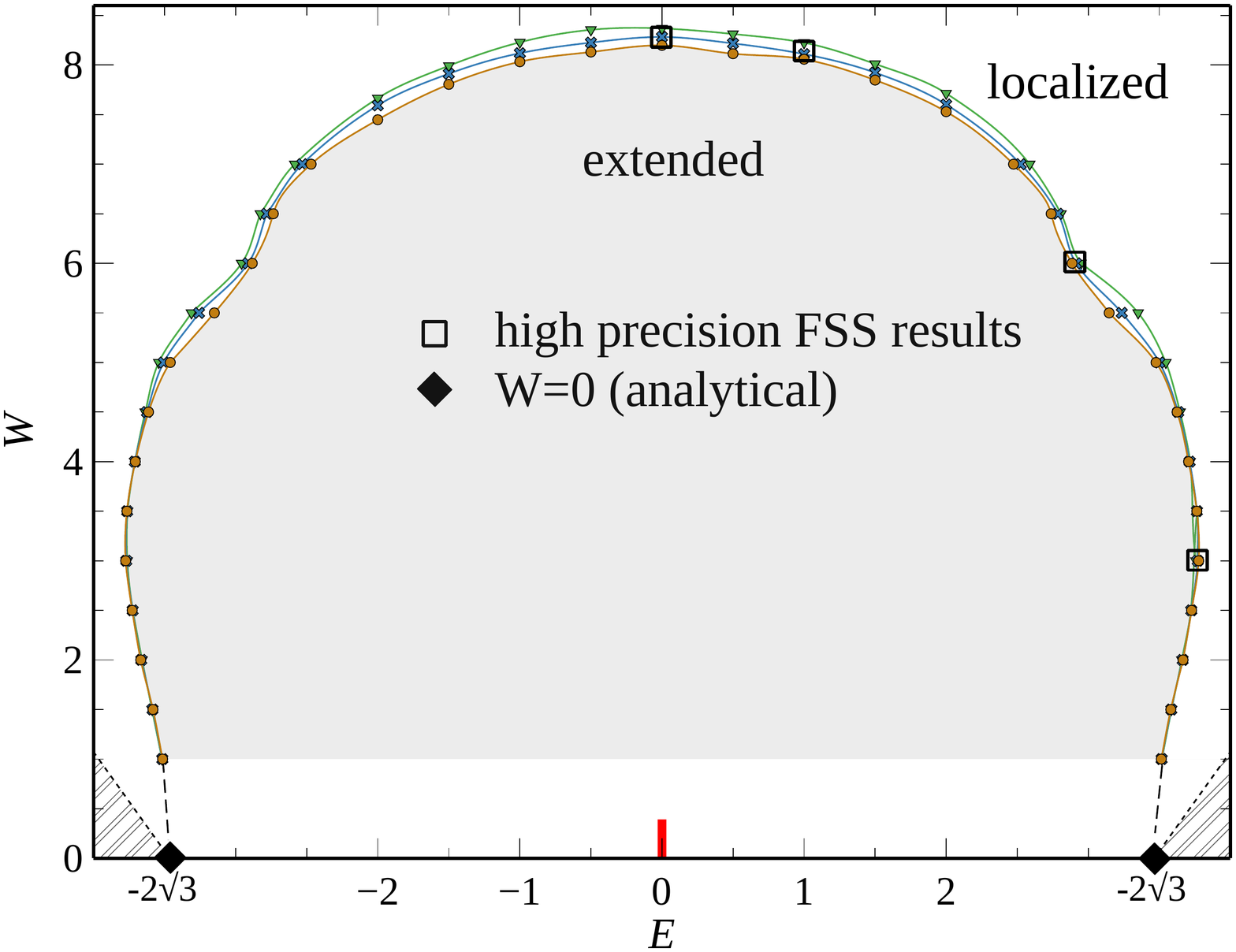}
    \caption{Phase diagram for $\mathcal{L}_3(1)$. The three solid and colored lines represent the approximate location of the phase boundary estimated from small $M$, i.e.\ the dark yellow line is constructed by widths $M=6$ and $M=8$, the blue line by $M=6$ and $M=10$ and the green line by $M=8$ and $M=10$. 
    The solid squares ($\square$) denote high-precision estimates from FSS for large $M$.
    The shaded area in the center contains extended states while states outside the phase boundary are localized. The dashed lines on both sides are guides-to-the-eye for the expected continuation of the phase boundary for $W<1$. 
    The red short vertical line at $E=0$ represents the position of the two-degenerate flat bands.
    The diamonds ($\blacklozenge$) denote the band edges for $W=0$, i.e.\ $E_\mathrm{min}=-2\sqrt{3}$ and $E_\mathrm{max}=2\sqrt{3}$. The dotted lines are the theoretical band edges $\pm \left(|E_\mathrm{min}| + -W/2\right)$ and the forbidden areas below those band edges have been filled by lines. 
    }
    \label{Fig:PhaseDiagramFor31}
\end{figure}

Fig.\ \ref{Fig:PhaseDiagramFor31} shows the energy-disorder phase diagram for $\mathcal{L}_3(1)$. 
The phase diagram was determined from the scaling behaviour of the $\Lambda(E,W)$ for small system sizes $M=6$, $M=8$ and $M=10$ with TMM error $\leq 0.1\% $ \cite{EILMES2008}. Data for $W<1$ fluctuates too much to give useful results and hence has been omitted from the figure. Clearly, the phase diagram is qualitatively similar to the phase diagram of the standard 3D Anderson model, although the band width and the critical disorder at $E=0$ are different. In particular, the critical disorder is reduced by about $50\%$ compared to the Anderson model. This is in agreement with the discussion in section \ref{sec:intro}. Close to the band edges for small $W\leq 4$ we also see a small re-entrant region as is also found in the 3D Anderson model. However, the shoulders that develop at $E\sim \pm 2.75$ and $W=6$ are a novel feature. The DOS at such strong disorder does not exhibit clearly any similar signatures.

\begin{figure*}[tbh]
    \centering
    (a)\includegraphics[width=0.98\columnwidth]{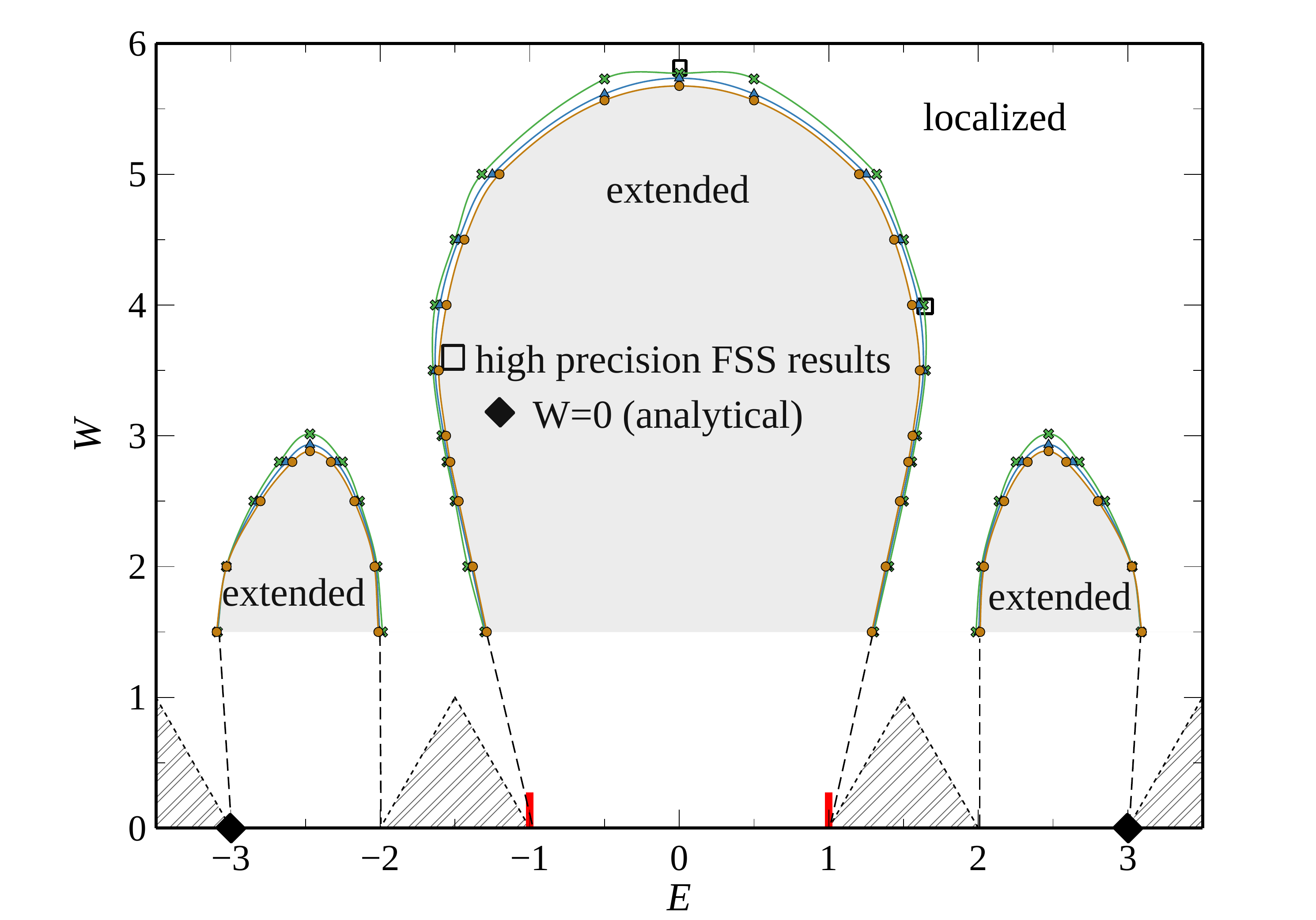}
    (b)\includegraphics[width=0.98\columnwidth]{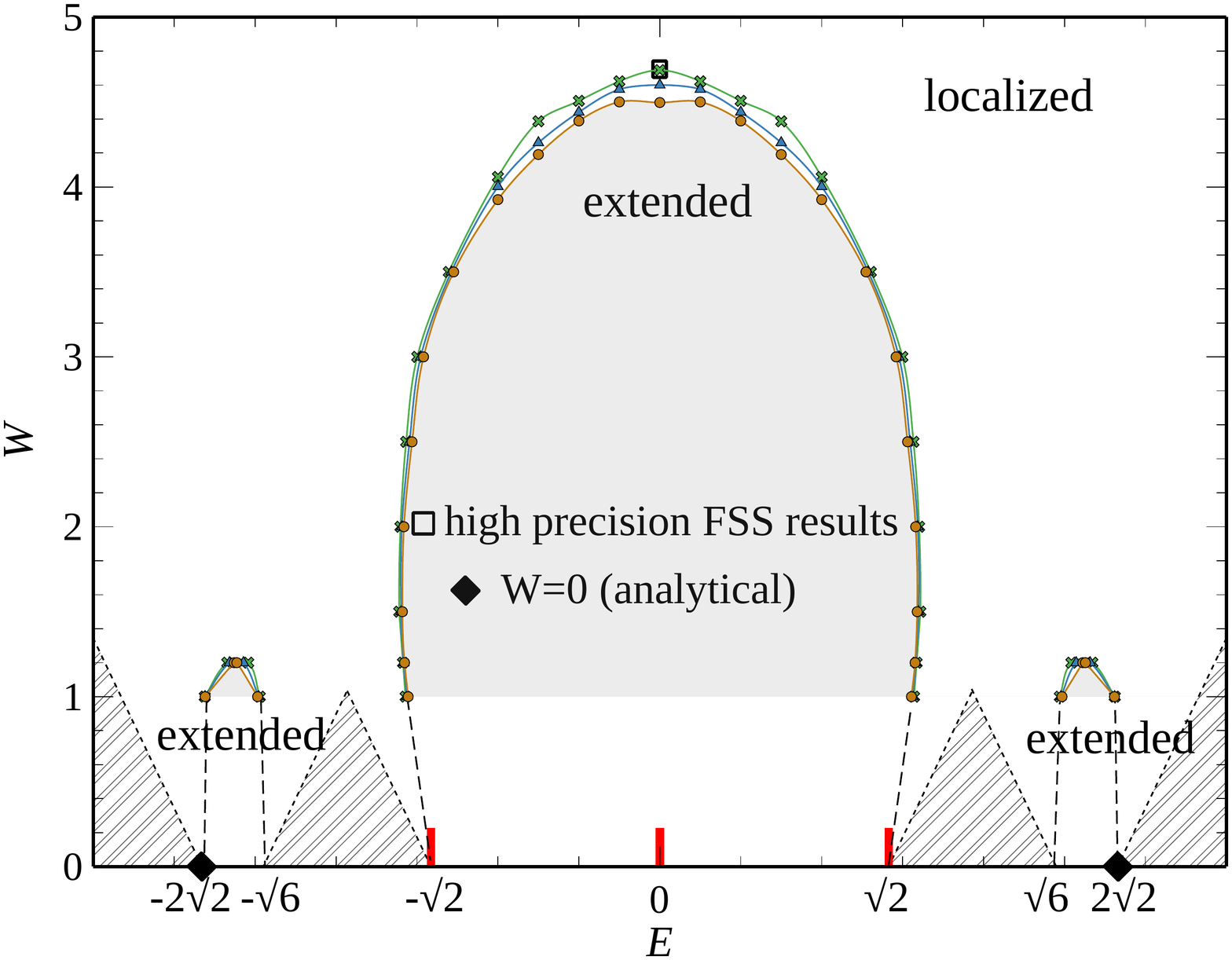}
    \caption{Phase diagrams for (a) $\mathcal{L}_{3}(2)$ and (b) $\mathcal{L}_{3}(3)$ lattices. The symbols, lines and colors are as in Fig.\ \ref{Fig:PhaseDiagramFor31}, i.e.\ representing small $M$ estimates with $M=6, 8$ and $10$. The solid squares ($\square$) denote high-precision FSS results from $\Lambda_M$ with an TMM error $\leq 0.1\%$ for width $M \leq 16$ and $\leq 0.2\%$ for width $M=18$. 
    The diamonds ($\blacklozenge$) denote the maximal band edges from $W=0$ at $\pm 3$ for $\mathcal{L}_{3}(2)$ and $\pm 2\sqrt{2}$ for $\mathcal{L}_{3}(3)$.
    }
    \label{fig:PhaseDiagramFor32and33}
\end{figure*}

For $ \mathcal{L}_3(2)$ and $ \mathcal{L}_3(3)$, we show the phase diagrams in Fig.\ \ref{fig:PhaseDiagramFor32and33}, determined with TMM errors of $\leq 0.2\% $ and with the same system sizes as for $\mathcal{L}_3(1)$.
As before, small disorder results have to be excluded. Our numerical results support, as for $\mathcal{L}_3(1)$, a mirror symmetry at $E=0$ and the results as shown in Fig.\ \ref{fig:PhaseDiagramFor32and33} have been explicitly symmetrized. For both $\mathcal{L}_3(2)$ and $ \mathcal{L}_3(3)$, the phase boundaries of the central dispersive band support a reentrant behaviour, although this is less so for $\mathcal{L}_3(3)$.

The obvious difference between the phase diagrams of $ \mathcal{L}_3(1)$, $ \mathcal{L}_3(2)$ and $ \mathcal{L}_3(3)$ is that the extended region for $ \mathcal{L}_3(1)$ lattice is simply connected, while for $ \mathcal{L}_3(2)$ and $ \mathcal{L}_3(3)$ it is disjoint. This difference can be attributed to the presence of the energy gaps in $ \mathcal{L}_3(2)$ and $ \mathcal{L}_3(3)$ as in Fig.\ \ref{fig:EnergyStructure}. Let us denote, as in the cubic Anderson model, a critical disorder $W_c$ as the disorder value at the transition point from extended to localized behaviour at energy $E=0$. Then we see that the critical disorders are  $W_c\sim 16.530$ for the cubic lattice \cite{Rodriguez2011MultifractalTransition}, $\sim 8.6$ for $\mathcal{L}_3(1)$, $\sim 5.9$ for $\mathcal{L}_3(2)$ and $\sim 4.8$ for $\mathcal{L}_3(3)$. Hence as expected, in the Lieb lattices the last extended states vanish already at much weaker disorders and the trend becomes stronger with increasing $n$ in each successive $\mathcal{L}_3(n)$.

\subsection{\label{sec:l31-23}High-precision determination of critical properties for the Lieb models}

\subsubsection{\label{sec:l31-3}Model $ \mathcal{L}_3(1)$}

In order to determine the critical properties at the phase boundaries for the Lieb models, we have to go to larger system size for a reliable FSS. In all cases, the results are collected up to $M=20$ and with TMM convergence errors $\leq 0.1\%$. Using the phase diagram as in Fig.\ \ref{Fig:PhaseDiagramFor31} as a rough guide, we pick out 4 points of special interest, namely, two transitions as a function of $W$ at the band centre at constant $E=0$ and outside the band centre at $E=1$. Furthermore, we also study two transitions as function of $E$ corresponding to the point marking the reentrant behaviour at constant $W=3$ and the kink in the phase boundary at constant $W=6$. 
In Fig.\ \ref{fig:3D_E0000,E0100,D0300,D0600}, we show the $\Lambda_M(E,W)$ data, the resulting scaling curves and the variation of the scaling parameter $\xi$ for typical examples of FSS results.
\begin{figure*}[tbh]
    \centering
    (a)\includegraphics[width=0.98\columnwidth]{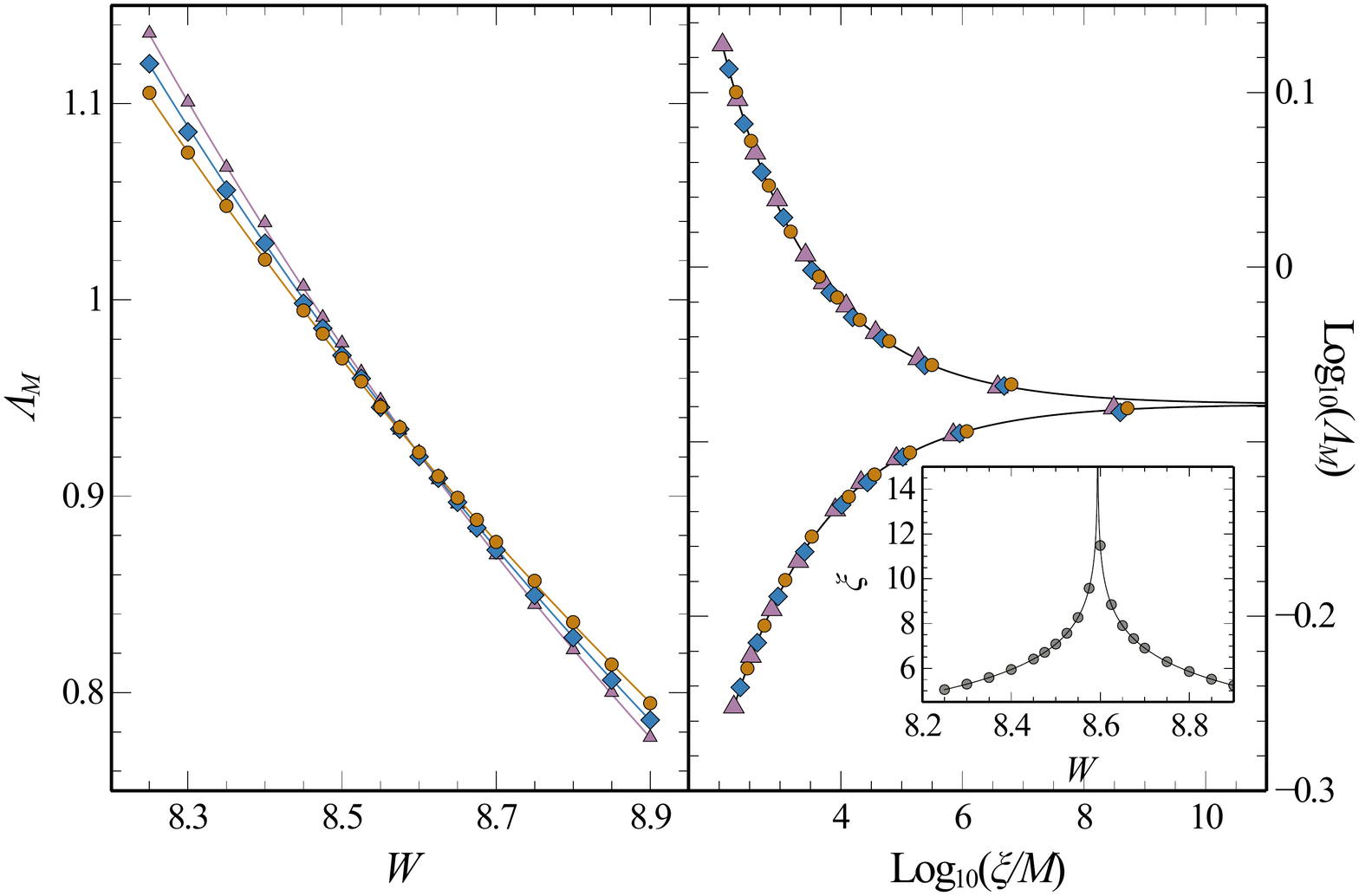}
    (b)\includegraphics[width=0.98\columnwidth]{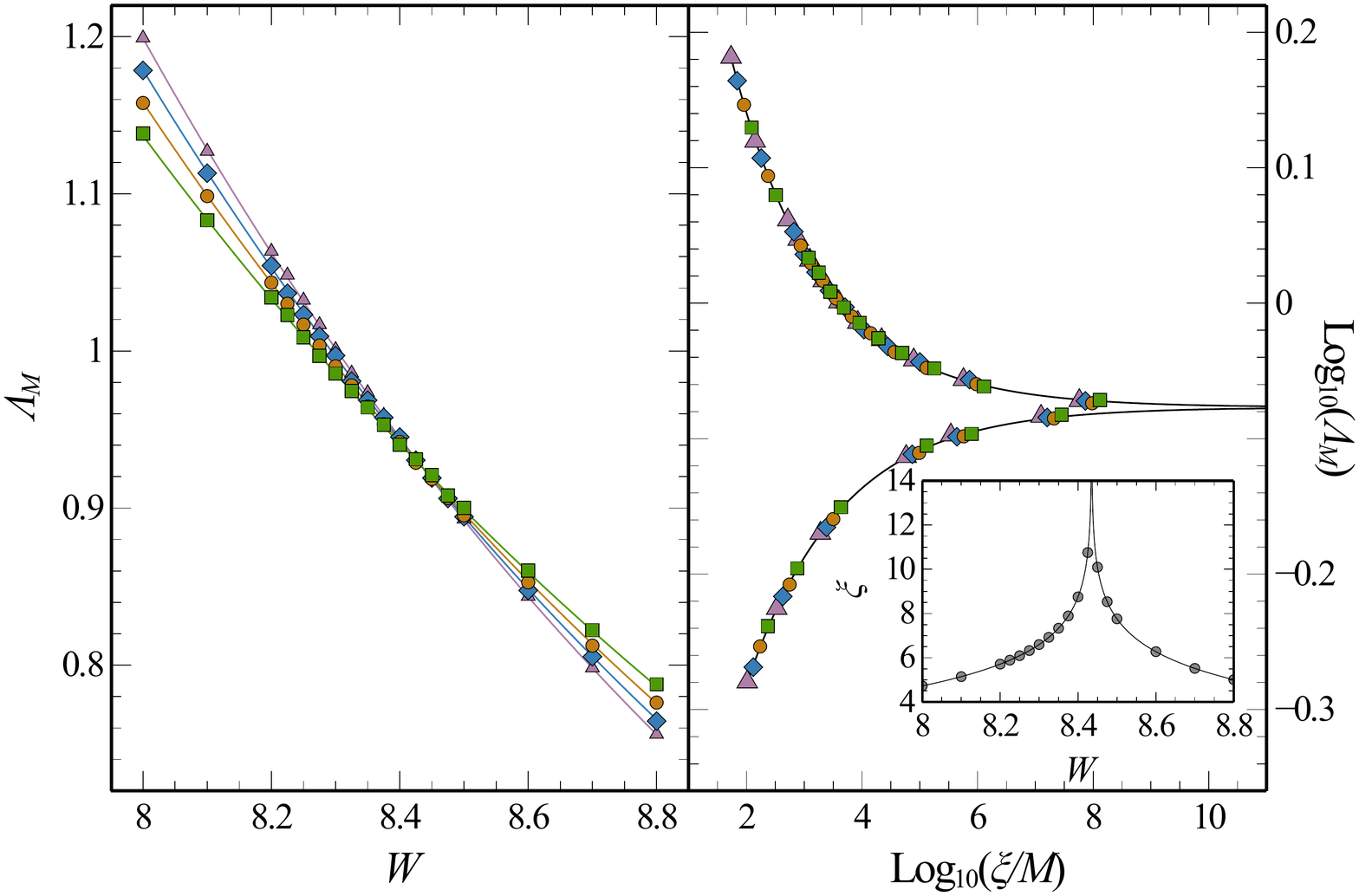}
    (c)\includegraphics[width=0.98\columnwidth]{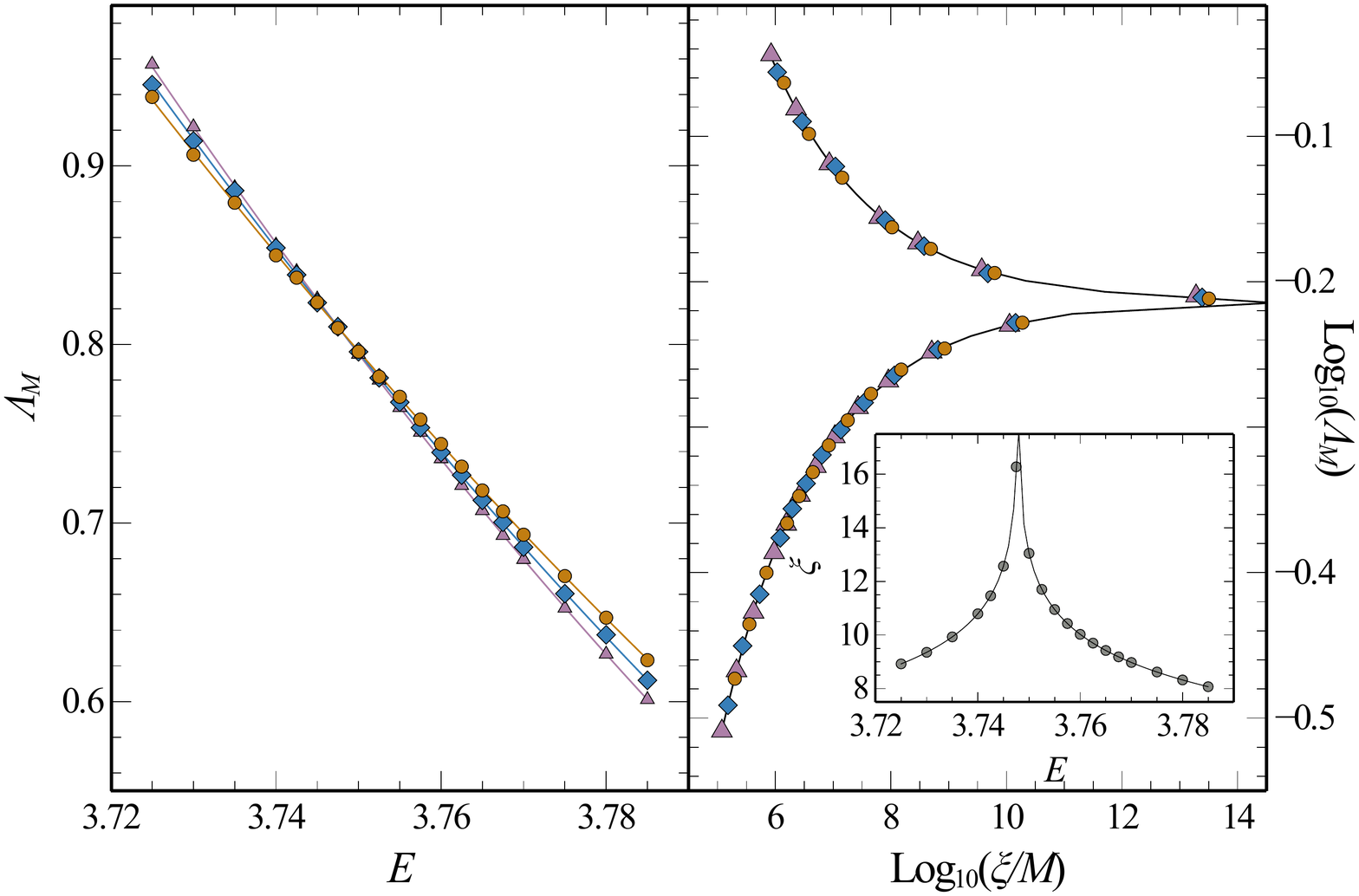}
    (d)\includegraphics[width=0.98\columnwidth]{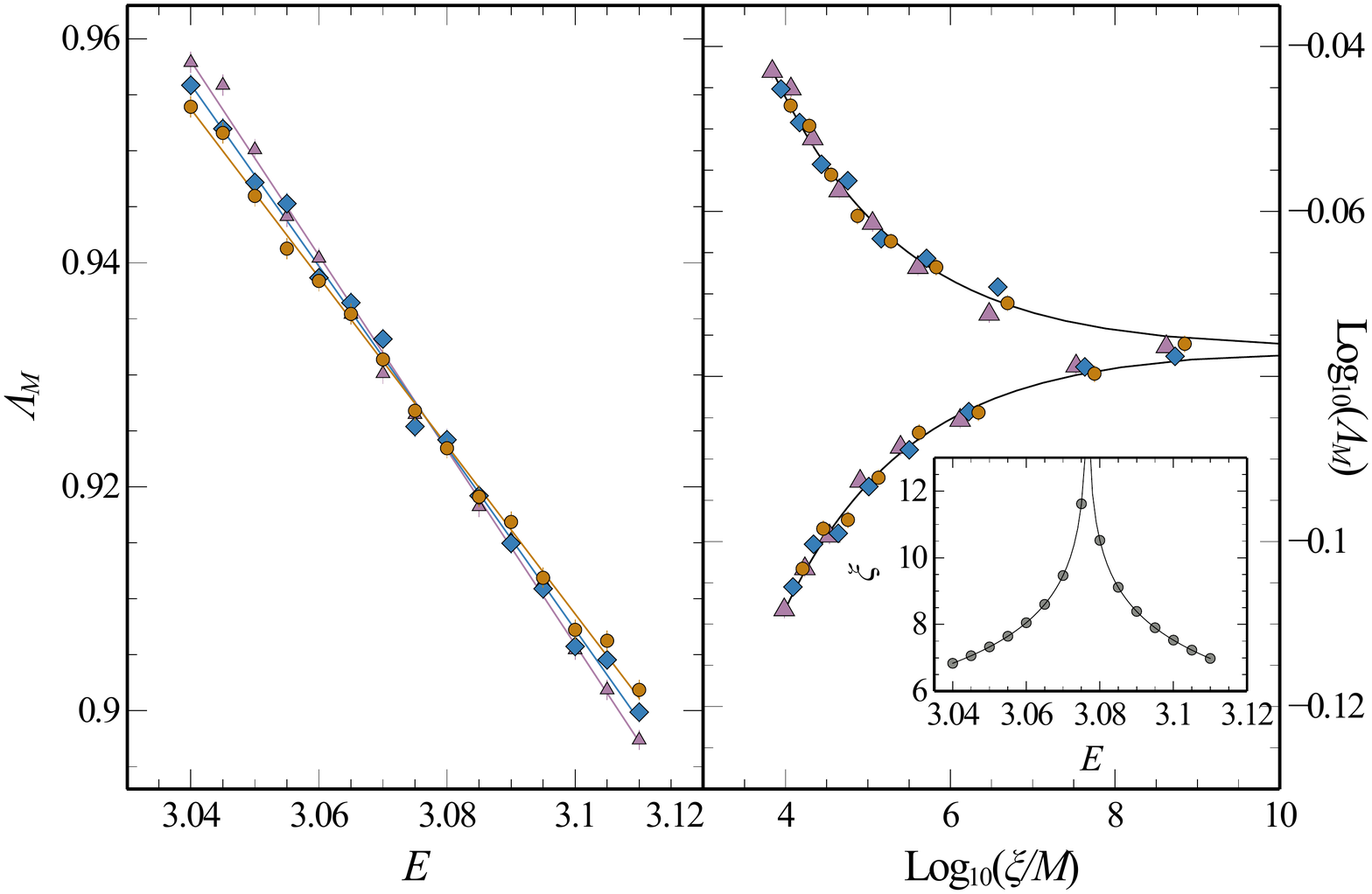}
     \caption{(a) FSS of the localization lengths for $\mathcal{L}_3(1)$ with $E = 0$, (b) $E = 1$, (c) $W = 3$, and (d) $W = 6$. System sizes $M$ are 14 ($\Box$), 16 ($\bigcirc$), 18 ($\Diamond$), 20 ($\bigtriangleup$). 
     The left half in each panel denotes a plot of $\Lambda_M$ versus disorder $W$ or energy $E$, the solid lines are fits to the data acquired by Eqs.\ \eqref{eq20}--\eqref{eq2} with (a+b) $n_r = 3$, $m_r = 1$, (c) $n_r = 2$, $m_r = 1$ and (d) $n_r = 1$, $m_r = 1$. 
     The right half in each panel shows the scaling function $F$ (solid line) and the scaled data points with the same $n_r$ and $m_r$ as in the corresponding left half while each inset gives the scaling parameter $\xi$ as a function of disorder strength $W$, in (a) and (b), or energy $E$ in (c) and (d).
     The parameters of the fits are shown in detail in Table \ref{table:critical parameters}.} 
      \label{fig:3D_E0000,E0100,D0300,D0600} 
\end{figure*}

In Table \ref{table:critical parameters} we present fits for all $4$ cases shown in Fig.\ \ref{fig:3D_E0000,E0100,D0300,D0600} with higher expansion coefficients  $n_r$ and $m_r$ that show that our results are stable with respect to an increase in an expansion parameter. 
\begin{table*}{}
\centering
\setlength{\tabcolsep}{3.25mm}{ 
\begin{tabular}{cccccccccc}
\hline \hline \noalign{\smallskip}
\multicolumn{10}{c}{$\mathcal{L}_3(1)$}\\
    $\Delta M$&   $E$&  $\delta W$&   $n_{r}$& $m_{r}$&    $W_{c}$&  CI($W_c$)&                            $\nu$&  CI($\nu$)&                       $p$        \\ 
    16-20&          0&    8.25-8.9&       3& 1&            $\textbf{\emph{8.594}}$& $[8.585,8.604]$&     $\textbf{\emph{1.57}}$& $[1.49,1.65]$&  $0.15$     \\ 
    16-20&          0&    8.25-8.9&       2& 2&            $8.598$&  $[8.586,8.610]$&                    $1.55$& $[1.46,1.63]$&                  $0.08$     \\
    16-20&          0&    8.25-8.9&       3& 2&            $8.595$&  $[8.582,8.607]$&                    $1.57$& $[1.48,1.66]$&                  $0.13$      \\
    Averages:&       &            &        &  &            $8.596(4)$&              &                    $1.56(3)$&                                          \\
\\
    $\Delta M$&   $E$&  $\delta W$&   $n_{r}$& $m_{r}$&     $W_{c}$&  CI($W_c$)&                        $\nu$&  CI($\nu$)&                       $p$       \\  
    14-20&          1&     8.0-8.8&        3& 1&            $\textbf{\emph{8.435}}$& $[8.429,8.441]$&   $\textbf{\emph{1.60}}$& $[1.54,1.65]$&   $0.18$     \\ 
    14-20&          1&     8.0-8.8&        2& 2&            $8.439$& $[8.432,8.447]$&                   $1.57$& $[1.53,1.62]$&                   $0.19$     \\
    14-20&          1&     8.0-8.8&        2& 3&            $8.438$& $[8.431,8.446]$&                   $1.57$& $[1.53,1.62]$&                   $0.21$      \\
    Averages:&       &            &         &  &            $8.437(3)$&             &                   $1.58(2)$&                                            \\    
\\
    $\Delta M$&  $W$&  $\delta E$&    $n_{r}$& $m_{r}$&       $E_{c}$&  CI($E_c$)&                         $\nu$&  CI($\nu$)&                      $p$      \\ 
    16-20&      3&    3.725-3.785&          2& 1&               $\textbf{\emph{3.748}}$& $[3.747,3.749]$&    $\textbf{\emph{1.75}}$& $[1.68,1.82]$&  $0.88$    \\ 
    16-20&      3&    3.725-3.785&          2& 2&               $3.748$& $[3.747,3.749]$&                    $1.76$& $[1.67,1.84]$&                  $0.86$    \\
    16-20&      3&    3.725-3.785&          3& 1&               $3.748$& $[3.747,3.749]$&                    $1.75$& $[1.68,1.82]$&                  $0.86$     \\
    Averages:&   &               &           &  &               $3.748(1)$&             &                    $1.75(3)$&           &                             \\
\\
    $\Delta M$&    $W$&   $\delta E$&   $n_{r}$& $m_{r}$&   $E_{c}$&  CI($E_c$)&                       $\nu$&  CI($\nu$)&                      $p$    \\ 
    16-20&           6&    3.04-3.11&       1& 1&           $\textbf{\emph{3.077}}$& $[3.070,3.083]$&  $\textbf{\emph{1.54}}$& $[1.08,2.01]$&  $0.14$  \\ 
    16-20&           6&    3.04-3.11&       2& 1&           $3.076$& $[3.069,3.082]$&                  $1.54$& $[1.09,1.99]$&                  $0.24$  \\
    16-20&           6&    3.04-3.11&       2& 2&           $3.077$& $[3.069,3.084]$&                  $1.54$& $[1.07,2.00]$&                  $0.21$  \\
    Averages:&        &             &        &  &           $3.077(3)$&             &                  $1.54(14)$&                                     \\  
\\[0.5ex]\hline\noalign{\smallskip}
\multicolumn{10}{c}{$\mathcal{L}_3(2)$}\\
    $\Delta M$ &    $E$&     $\delta W$&      $n_{r}$& $m_{r}$&      $W_{c}$&  CI($W_c$)&                          $\nu$&  CI($\nu$)&                       $p$   \\ 
    12,14,18  &     0  &     5.85-6.05 &          2& 2&               $\textbf{\emph{5.964}}$& $[5.958,5.969]$&    $\textbf{\emph{1.75}}$& $[1.57,1.92]$&   $0.08$ \\
    12,14,18  &     0  &     5.85-6.05 &          2& 3&               $5.965$& $[5.959,5.970]$&                    $1.70$& $[1.51,1.89]$&                   $0.08$ \\ 
    12,14,18  &     0  &     5.85-6.05 &          3& 2&               $5.963$& $[5.956,5.971]$&                    $1.75$& $[1.57,1.92]$&                   $0.07$ \\
    Averages: &        &               &           &  &               $5.964(3)$&              &                    $1.73(6)$&                                       \\
\\
    $\Delta M$&     $W$&    $\delta E$&       $n_{r}$& $m_{r}$&  $E_{c}$&  CI($W_c$)&                         $\nu$&  CI($\nu$)&                        $p$         \\ 
    10,12,14  &     4&      1.6-1.8 &            2& 1&           $\textbf{\emph{1.704}}$& $[1.701,1.708]$&    $\textbf{\emph{1.55}}$& $[1.43,1.68]$&    $0.18$      \\
    10,12,14  &     4&      1.6-1.8 &            1& 3&           $1.705$& $[1.701,1.709]$&                    $1.56$& $[1.43,1.70]$&                    $0.1$       \\
    10,12,14  &     4&      1.6-1.8 &            2& 2&           $1.703$& $[1.700,1.707]$&                    $1.53$& $[1.40,1.66]$&                    $0.2$       \\
    Averages: &      &              &             &  &           $1.704(2)$&             &                    $1.55(5)$&                                             \\  
\\[0.5ex]\hline\noalign{\smallskip}
\multicolumn{10}{c}{$\mathcal{L}_3(3)$}\\[0.5ex]
    $\Delta M$&    $E$&      $\delta W$&      $n_{r}$& $m_{r}$&  $W_{c}$&  CI($W_c$)&                            $\nu$&  CI($\nu$)&                     $p$     \\ 
    12-18     &     0 &     4.7--4.875 &        2& 1&            $\textbf{\emph{4.79}}$& $[4.786,4.794]$&       $\textbf{\emph{1.63}}$& $[1.48,1.78]$& $0.49$  \\
    12-18     &     0 &     4.7--4.875 &        1& 2&            $4.791$& $[4.786,4.795]$&                       $1.63$& $[1.48,1.78]$&                 $0.47$   \\
    12-18     &     0 &     4.7--4.875 &        2& 2&            $4.791$& $[4.786,4.795]$&                       $1.63$& $[1.48,1.78]$&                 $0.47$   \\
    Averages:&        &                &         &  &            $4.790(2)$ &             &                      $1.63(5)$&                                      \\
 \hline\hline

\end{tabular}
\caption{Critical parameters of the MIT for $\mathcal{L}_3(n)$, $n=1, 2, 3$ and $4$. The columns denoting system width $M$, fixed $E$ (or $W$), range of $W$ (or $E$), expansion orders $n_{r}$, $m_{r}$ are listed as well as resulting critical disorders $W_c$ (or energies $E_c$), their 95$\%$ confidence intervals (CI), the critical exponent $\nu$, its CI, and the goodness of fit probability $p$. The averages contain the mean of the three preceding $W_c$ (or $E_c$) and $\nu$ values, with standard error of the mean in parentheses. The bold $W_c$, $E_c$ and $\nu$ values highlight the fits used as examples in Figs.\ \ref{fig:3D_E0000,E0100,D0300,D0600} and \ref{fig:L32-3_E0000_FSS_Diagram}.}
\label{table:critical parameters}}
\label{tab:L32-3_E0000_FSS}
\end{table*}
We have also checked that they are stable with respect to slight changes in the choice of parameter intervals $\delta W$ and $\delta E$ for fixed energy and fixed disorder transitions, respectively. The reader will have noticed, however, that the accuracy of the data is not good enough to reliably fit irrelevant scaling contributions and hence the results in Table \ref{table:critical parameters} are all for $n_i=m_i=0$ although we have indeed performed our FSS allowing for these additional parameters. Furthermore, one can see in Fig.\ \ref{fig:3D_E0000,E0100,D0300,D0600} that the accuracy of the TMM data becomes worse for the fixed disorder transitions at $W=3$ and especially $W=6$. The reason for this behaviour is in principle well understood since at the points, the DOS has an appreciable variation which leads to extra corrections not well captured in the FSS \cite{Cain1999PhaseHopping}. Usually, larger system sizes $M$ can reduce these variations but this is not possible here due to computational limitations.

\subsubsection{\label{sec:2-3}Models $\mathcal{L}_3(2)$ and $ \mathcal{L}_3(3)$}

\begin{figure}[tbh]
    \centering
    (a)\includegraphics[width=0.95\columnwidth]{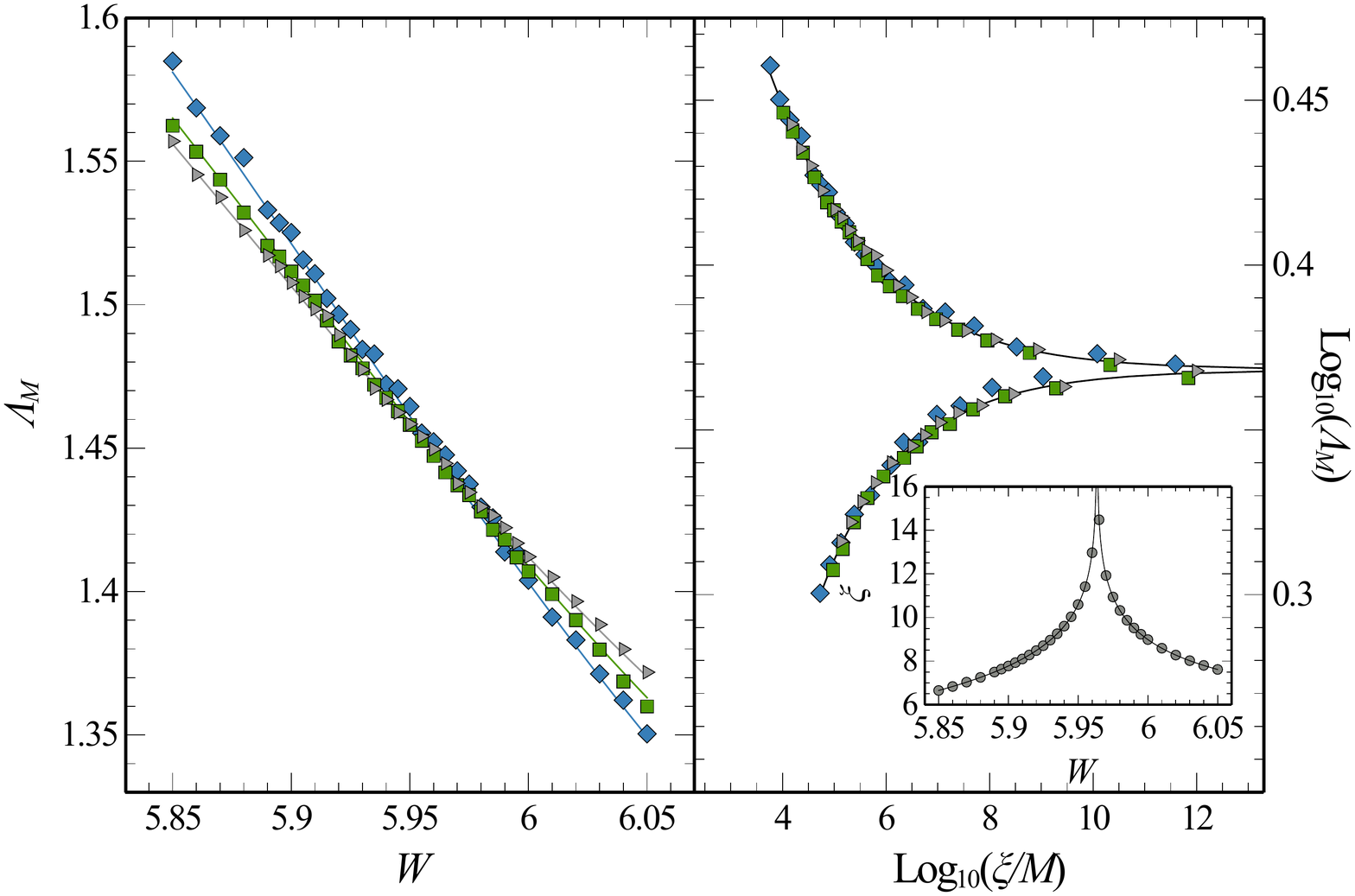}
    (b)\includegraphics[width=0.95\columnwidth]{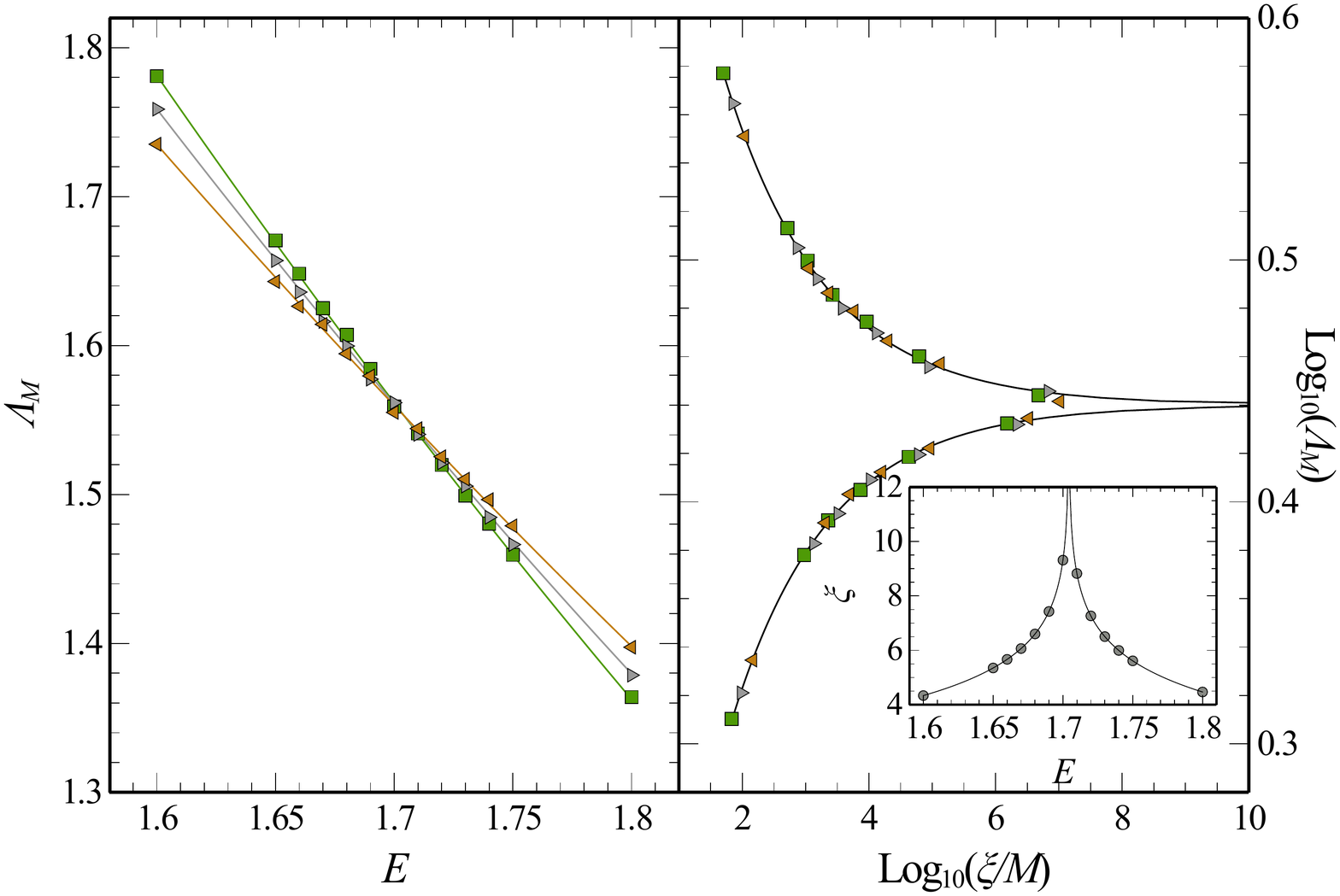}
    (c)\includegraphics[width=0.95\columnwidth]{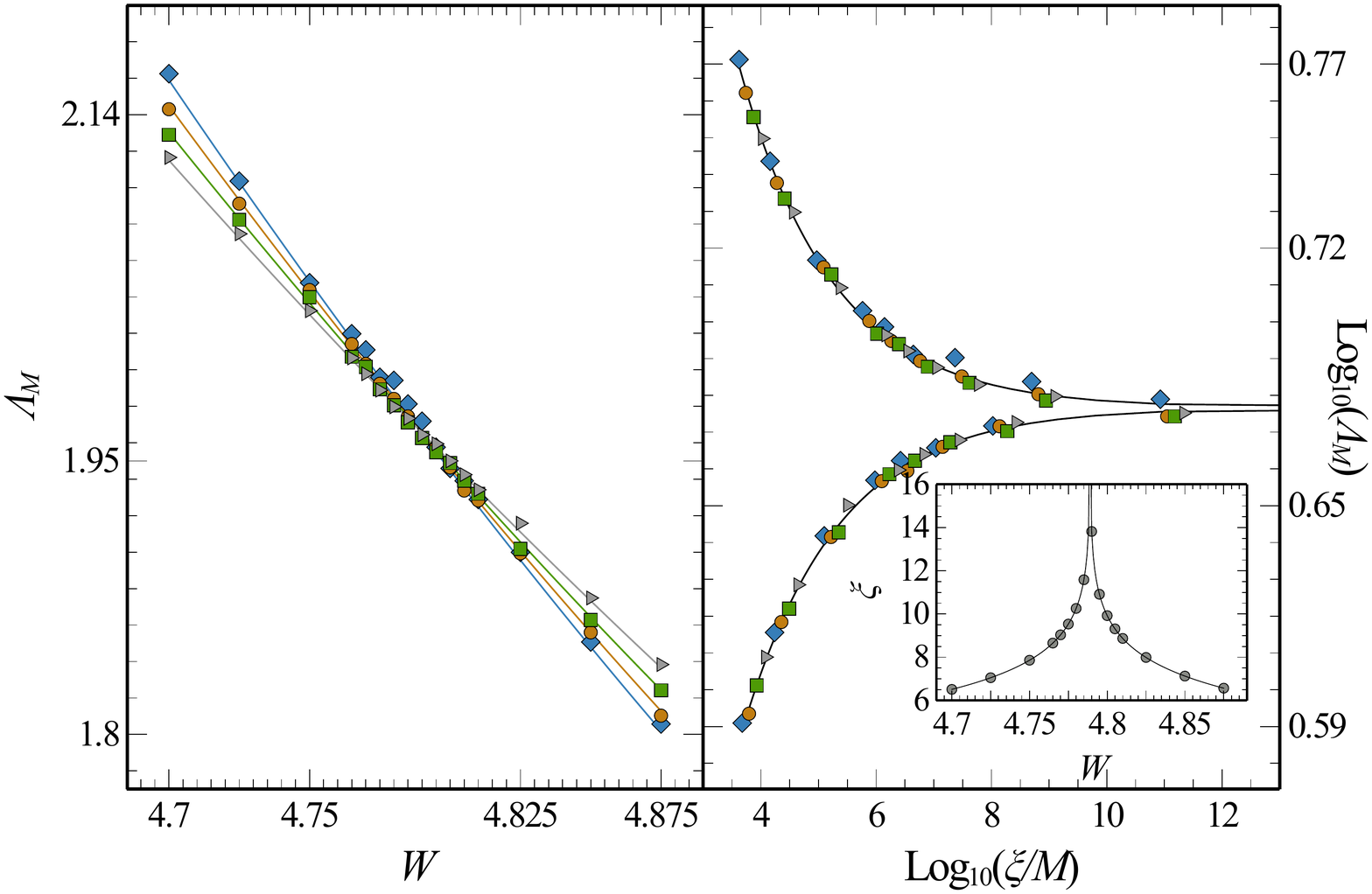}
    \caption{
     FSS of the localization lengths for (a) $\mathcal{L}_3(1)$ at $E = 0$ and (b) $W = 4$ as well as for (c) $\mathcal{L}_3(3)$ at $E = 0$. System sizes $M$ are 10 ($\triangleleft$), 12 ($\triangleright$), 14 ($\Box$), 16 ($\bigcirc$), 18 ($\Diamond$).
     The arrangement in each panel is as in Fig.\ \ref{fig:3D_E0000,E0100,D0300,D0600}, i.e.\ scaling curves (solid lines) and scaled $\Lambda_M$ data (symbols) in the left half of each panel, scaling curve $F$ (lines) with scaled data (symbols) in the right half and $\xi$ in the inset.
     The chosen expansion coefficients are (a) $n_r = 2$, $m_r = 2$, (b) $n_r = 2$, $m_r = 1$ and (c) $n_r = 2$, $m_r = 1$ as highlighted in Table \ref{table:critical parameters}.
     }
    \label{fig:L32-3_E0000_FSS_Diagram}
\end{figure}

We follow a similar strategy as in the previous section in order to finite-size scale the localization lengths for $\mathcal{L}_3(2)$ and $\mathcal{L}_3(3)$. The TMM convergence errors were chosen as $\leq 0.1\%$  up to $M=16$ and, due to the increased complexity of these models, as $\leq 0.2\%$ for the largest system size with $M=18$.
Fig.\ \ref{fig:L32-3_E0000_FSS_Diagram}(a) shows $\Lambda_M(E=0,W)$ and the scaling curve for $\mathcal{L}_3(2)$ at energy $E=0$ with $n_r=3, m_r=3$. From the panel with the $\Lambda_M(E=0,W)$ data, it is very hard to observe a clear crossing at $W_c$. The situation improves for $\Lambda_M(E,W=4)$ in Fig.\ \ref{fig:L32-3_E0000_FSS_Diagram}(b)
which exhibits a clear crossing of $\Lambda_M$ around $E_c \sim 1.70$. For $\mathcal{L}_3(3)$ shown in Fig.\ \ref{fig:L32-3_E0000_FSS_Diagram}(c) the crossing for $\Lambda_M(E=0,W)$ is again somewhat less clear. Nevertheless, in all three cases, the FSS results produce stable and robust fits with estimates for $W_c$, $E_c$ and $\nu$ as shown in Table \ref{table:critical parameters}. As for $\mathcal{L}_3(1)$, the FSS fits  $\mathcal{L}_3(2)$ and  $\mathcal{L}_3(3)$ do not resolve potential irrelevant scaling corrections. 
%

\section{\label{sec:conclusions}Conclusions}

\begin{figure}[tbh]
    \centering
    \includegraphics[width=0.98\columnwidth]{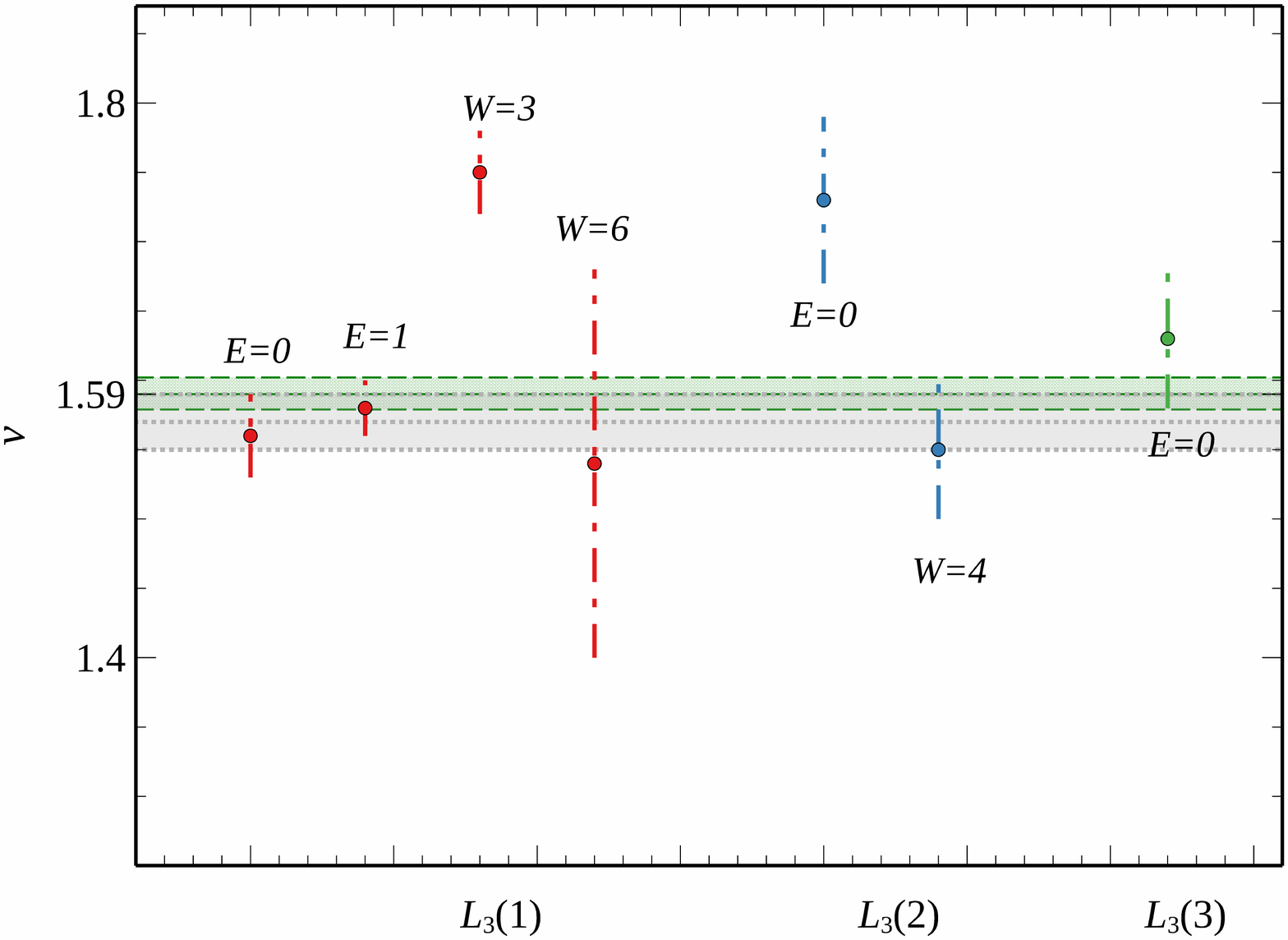}
    \caption{Variation of the averaged critical exponent $\nu$ corresponding to $\mathcal{L}_3(1)$ (red), $\mathcal{L}_3(2)$ (blue) and $\mathcal{L}_3(3)$ (green) for the seven averages from Table \ref{table:critical parameters}. The green horizontal dash lines indicate $\nu=1.590 (1.579,1.602)$ via FSS of wave functions in the 3D Anderson model \cite{Rodriguez2011MultifractalTransition} and the green shadow area denotes its error range. The $\nu=1.57(2)$ value, indicated by grey dotted lines with grey shadow area denoting its error bar, is from TMM results \cite{Slevin1999b}.}
    \label{fig:my_label}
\end{figure}

There are two ways to understand the Lieb lattices as originating from the normal simple cubic lattices: (i) as shown in Fig.\ \ref{Fig:Lieb3D_Graph}, one can view the $\mathcal{L}_3(n)$ lattices as a cubic lattice with additionally added sites between the vertices of the cube, effectively allowing for additional back-scattering and interference along the original site-to-site connections and hence potentially leading to more localization. On the other hand, one might argue that (ii) the $\mathcal{L}_3(n)$ lattices can be constructed by deleting sites from a cubic lattice, for example a central site in Fig.\ \ref{Fig:Lieb3D_Graph}(a) and the 6 face-centered sites. In this view, the decrease of possible transport channels should give rise to stronger effective localization. Both constructions lead to the same predictions and agree with what we find here, namely, the localization properties in all $\mathcal{L}_3(n)$ lattices show an increased localization with respect to the cubic Anderson lattice \emph{and} become stronger when $n$ increases. This is, for example, clear from looking at the behaviour of $W_c(n)$ in Table \ref{table:critical parameters}. 
It is instructive to study the behaviour as $n\rightarrow \infty$. From Fig.\ \ref{fig:EnergyStructure}, we see that the overall band width decreases as $n$ increases. At the same time, the number of flat bands increases and the extremal energy of these bands extends as well towards $|E|=2$. Hence for very large $n$, where $\mathcal{L}_3(n)$ is simply a $2^3$ renormalized lattice, but $n$ renormalized sites apart, with proliferating flat bands.
%
Our results for the critical exponent then suggest that as $n$ increases and the dispersive bands become smaller, the critical properties in each band still retain the universality of the 3D Anderson transition --- at least up to $n=3$ that we have been able to compute (cp.\ Fig.\ \ref{fig:my_label}. This is in good agreement with previous results in loosely coupled planes of Anderson models in which the universal 3D behaviour was also retained \cite{Milde2000CriticalSystems}. However, for loosely coupled planes, the MIT was retained even for small interplane coupling --- a truly 2D localization behavior only emerged when the interplace coupling was zero. The point of view of this work is different, i.e.\ the change from 3D dispersive bands with an MIT to a solely 1D system without MIT is not a continuous change, but rather an eventual replacement and shrinking of dispersive bands by a proliferation of flat bands as $n$ grows.

\section*{\label{sec:acknow}Acknowledgments}
We wish to acknowledge the National Natural Science Foundation of China (Grant No.\ 11874316), the Program for Changjiang Scholars and Innovative Research Team in University (Grant No.\ IRT13093), and the Furong Scholar Program of Hunan Provincial Government (R.A.R.) for financial support. This work also received funding by the CY Initiative of Excellence (grant "Investissements d'Avenir" ANR-16-IDEX-0008) and developed during R.A.R.'s stay at the CY Advanced Studies, whose support is gratefully acknowledged. We thank Warwick's Scientific Computing Research Technology Platform for computing time and support. UK research data statement: Data accompanying this publication are available from the corresponding authors.

\appendix
\section{\label{Appendix}Dispersions}
For completeness, we here include the dispersion relations shown in Fig.\ \ref{fig:EnergyStructure}. For $\mathcal{L}_3(2)$, we have 
\begin{subequations}
\begin{align}
    &E_{1,2}= 1, \quad \hfill E_{3,4}= -1, & E_{5}= \rho_+ + \rho_-,\\
    &E_{6}= \omega \rho_+ + \omega^2 \rho_-,
    &E_{7}= \omega \rho_- + \omega^2 \rho_+,
\end{align}
\end{subequations} 
where $\omega=\frac{-1+i\sqrt{3}}{2}$, $\rho_{\pm}= \sqrt[3]{-\frac{q(\vec{k})}{2} \pm \sqrt{\left(\frac{q(\vec{k})}{2}\right)^2 - \left(\frac{7}{3}\right)^3}}$ and $q(\vec{k})=2\left(\cos k_{x} + \cos k_{y} + \cos k_{z}\right)$.
For $\mathcal{L}_3(3)$, we find  
\begin{subequations}
\begin{align}
    &E_{1,2}= \sqrt{2},\quad
     E_{3,4}= -\sqrt{2},\quad
     E_{5,6}= 0,\\
    &E_{7,8,9,10}= \pm \sqrt{4 \pm \sqrt{10 + q(\vec{k})}}.
\end{align}
\end{subequations}
Last, for $\mathcal{L}_3(4)$, the four doubly degenerate flat bands are given as 
\begin{subequations}
\begin{equation}
    E_{1,2,3,4,5,6,7,8}= \frac{1}{2}\left(\pm 1 \pm  \sqrt{5}\right),
\end{equation}
and the remaining five dispersive bands are the solutions of the $5$th order equation 
\begin{equation}
E^5 - 9E^3 + 13E - q(\vec{k}) = 0.
\end{equation}
\end{subequations}


\begin{thebibliography}{50}
\bibitem{psiAD} Strictly speaking, the labels A and D are not needed to label $\Psi^{A,D}_z$, but we retain them here for the readers convenience.
\end{thebibliography}

\end{document}